# A Multiphase Absorber Containing O VI and Broad H I Directly Tracing $10^6$ K Plasma at Low-Redshift Toward HE 0153-4520[1]

B. D. Savage[2], A. Narayanan[2,3], N. Lehner[4], B. P. Wakker[2]

ABSTRACT

Observations of the QSO HE 0153-4520 ($z_{em}$ = 0.450) with the Cosmic Origins Spectrograph (COS) from 1134 to 1796 Å with a resolution of ~17 km $s^{-1}$ and signal-to-noise (S/N) per resolution element of 20 to 40 are used to study a multi-phase partial Lyman Limit system (LLS) at z = 0.22601 tracing both cool and hot gas. FUSE observations of the Lyman limit break yield log N(H I) = 16.61(0.12, -0.17) The observed UV absorption lines of H I λλ 1216 to 926, C III, C II, N III, N II, Si III, and Si II imply the existence of cool photoionized gas in the LLS with log U = -2.8±0.1 and log N(H) = 19.35±0.18, log $n_H$ = -2.9±0.2, log T = 4.27±0.02, log (P/k) = 1.75±0.17, and log L(kpc) = 0.70±0.25. The abundances are [X/H] = -0.8 (+0.3, -0.2) for N, Si and C but the result is sensitive to the assumed shape of the ionizing background radiation field. The multi-phase system has strong O VI and associated broad Ly α absorption (BLA) with log N(O VI) = 14.21±0.02, b(O VI) = 37±1 km $s^{-1}$, log N(H I) = 13.70(+0.05,-0.08), b(H I)=140 (+14, -16) km $s^{-1}$ and b(H I)/b(O VI) = 3.9±0.4. The O VI does not arise in the cool photoionized gas of the LLS. The O VI and BLA imply the direct detection of thermally broadened absorption by hot gas with log T = 6.07 (+0.09, -0.12), [O/H] = -0.28 (+0.09, -0.08) , and log N(H) = 20.41 (+0.13, -0.17). The absorber probably occurs in the circumgalactic environment (halo) of a foreground galaxy. The very high sensitivity of UV absorption line spectroscopy to O VI and thermally broadened H I Ly α now permits the direct detection of hot gas with log T > 6, a determination of the abundance of O/H, and log N(H) in low-redshift IGM absorption line systems.

*Short title: An O VI Absorber Tracing $10^6$ K Gas*

---

[1] Based on observations obtained with the NASA/ESA Hubble Space Telescope, which is operated by the Association of Universities for Research in Astronomy, Inc. under NASA contract NAS5-26555, and the NASA-CNES/ESA Far Ultraviolet Spectroscopic Explorer mission operated by Johns Hopkins University, supported by NASA contract NAS 05-32985.
[2] Department of Astronomy, University of Wisconsin-Madison, 475 N. Charter St., Madison, WI. 53706
[3] Indian Institute of Space Science and Technology, Thiruvananthapuram 695547, Kerala, India
[4] Department of Physics, University of Notre Dame, 225 Nieuwland Science Hall, Notre Dame, IN 46556

## 1. INTRODUCTION

At low redshift ~90% of the normal baryonic matter likely resides in the regions between galaxies with only ~10% found in luminous objects (stars and galaxies) (Fukugita & Peebles 2004). The cooler ~30% of the baryons with T ~ (2-4)x$10^4$ K have been traced by the Lyman $\alpha$ forest (Penton et al. 2004; Lehner et al. 2007; Danforth & Shull 2008) and some of the O VI lines (Tripp et al. 2008; Thom & Chen 2008). The warmer ~ 20 to 30% of the baryons with T ~ (1-10)x$10^5$ K have been traced by broad Ly$\alpha$ (Richter et al. 2006; Lehner et al. 2007; Danforth et al. 2010a), O VI absorption (Danforth & Shull 2008; Tripp et al. 2008) and Ne VIII absorption (Savage et al. 2005; Narayanan et al. 2009, 2011) although the baryonic content of these warm absorbers is very uncertain. The remaining ~40% of the baryons probably resides in hotter gas with T > $10^6$ K. The hot baryons can be studied at X-ray wavelengths through searches for O VII and O VIII and other X-ray lines (see the review of Bregman 2007). At UV wavelengths hot baryons can be studied via redshifted EUV absorption lines of Ne VIII, Mg X and Si XII, by broad but weak absorption by O VI (Savage et al. 2010) and by very broad H I Lyman $\alpha$ absorption (this paper).

It has proven easier to establish the probable existence of baryons in photoionized gas with T ~ (1-4)x$10^4$ K than in collisionally ionized gas at higher temperatures. This is because the neutral hydrogen that exists in photoionized gas produces absorption lines that are relatively strong and therefore allow good determinations of log N(H I) and the Doppler parameter b (km $s^{-1}$). If the photoionized gas also contains metal absorption lines, the decrease in the observed line widths as the mass of the absorbing ion increases can be used to simultaneously solve for the thermal and the non-thermal contributions to the line broadening and thereby estimate the temperature in the gas. This is done by assuming $b^2 = b_t^2 + b_{nt}^2$, with $b_t = (2kT/m)^{1/2} = 0.129\,(T/A)^{1/2}$ km $s^{-1}$ where b, $b_t$ and $b_{nt}$ are the observed, the thermal, and the non-thermal values of the Doppler parameter, m is the mass and A is the atomic mass number for the absorbing atom. In the equation above, the non-thermal broadening is assumed to have a Gaussian line shape and the different species are assumed to co-exist in a simple isothermal structure. These simplifying assumptions limit the reliability of the method. Another complication is the good alignment of absorption by photoionized H I and narrower O VI absorption could occur if the O VI traces collisionally ionized gas at the interface between the cooler photoionized gas and a hot exterior medium. Observations of O VI absorption in cool/hot gas interfaces in local interstellar medium reveal the O VI absorption is usually well aligned with absorption by the cool gas absorbers including O I and C II (Savage & Lehner 2006). Therefore, well aligned H I and O VI absorption with the difference in b values implying gas temperatures consistent with photoionization does not definitely prove photoionization is actually creating the O VI.

Determining whether the O VI is collisionally ionized is difficult because the H I associated with hotter collisionally ionized gas is often relatively weak, very broad and difficult to detect. Also, the O VI absorption may blend with cooler H I associated with photoionized gas in the same system. It therefore is not surprising that the first major surveys aimed at probing the physical properties of O VI absorption in the IGM by Tripp et al. (2008) and Thom & Chen (2008) revealed that ~25% of the O VI absorbers could be explained by photoionization in gas with T ~ (1-4)x$10^4$ K. However, those surveys were unable to determine the relative amounts of photoionized to collisionally ionized gas because of the difficulties associated with finding good evidence that some of the O VI absorbers are tracing gas with $T > 10^5$ K. A number of investigators have taken the lack of good evidence for collisionally ionized O VI and the strong evidence for photoionized O VI and have concluded that most of the O VI in the IGM is photoionized. That conclusion has been reinforced by the theoretical simulations of Oppenheimer & Davé (2009) and Kang et al. (2005) who conclude that the vast majority of the O VI found in absorption systems is photoionized. However, the alternate simulations of Tepper-García et al. (2011) and Smith et al. (2010) imply that the majority of the O VI absorbers trace metal enriched collisionally ionized gas with log T = 5.3±0.5. Modifying the assumptions regarding the basic physics of the cooling of the gas in the more recent simulations has completely changed the outcome. Reliable observations are required to establish whether or not the basic physics employed in these very complex simulations is done correctly.

Tripp et al. (2008) reported results for 77 O VI components. However, temperature estimates were only given for 28 well aligned O VI and H I components. An examination of the 28 components in Table 7 of Tripp et al. (2008) reveals 20 components with log T < 4.7, 8 with log T > 4.7 and 4 components with log T > 5. It is unlikely for gas with temperatures as hot as log T > 4.7 or 50,000 K to be in photoionization equilibrium. This simple exercise then suggests that among the 28 well aligned O VI and H I components 71% have properties consistent with photoionization and 29% are inconsistent with photoionization. Therefore 20 of 77 or ~25% of the O VI components have properties consistent with photoionization in the Tripp et al. (2008) sample. However, there is very little information regarding the temperature or ionization processes operating in the 49 additional O VI components in the Tripp et al. survey that are not well aligned with the H I absorption. In these cases the observations provide strong evidence for the existence of multiple gas phases in the absorbers. The correct conclusion to draw from the observations is that the origin(s) of the ionization in ~75% of the O VI absorption line systems studied by Tripp et al. (2008) is not well understood. This is unfortunate because without understanding the ionization process it is not possible to determine the metallicity of the gas and to relate the measures of O VI column density to estimates of the total associated hydrogen column density.

For a number of O VI absorption systems analyzed recently, it has proven possible to show that collisional ionization in gas with $10^5 < T < 10^6$ K is likely the dominant ionization process. Table 1 lists 14 O VI absorbers where the evidence for collisional ionization in gas with log T > 5 is relatively strong. In three systems the detection of O VI and of Ne VIII implies collisionally ionized gas because Ne VIII is extremely difficult to produce by photoionization in the low redshift IGM. This is because a photoionized origin of Ne VIII by the general extragalactic radiation field requires a very low gas density and very large (>several Mpc) path lengths. Over such long path lengths, Hubble flow broadening would produce Ne VIII absorption lines many times broader than the observed line widths. In other cases, the collisional ionization origin of the O VI follows from the detection of O VI and the detection of associated weak but very broad H I absorption. As in the case of the analysis of the photoionized O VI systems, the difference in line width allows a determination of the thermal Doppler and non-thermal (turbulent) contribution to the absorption line broadening with the thermal contribution requiring log T > 5. Both of these techniques for finding collisionally ionized O VI are difficult to implement because the Ne VIII and broad H I lines are weak and only apparent in spectra with relatively high S/N. The high S/N is required not only to detect the broad H I but to also to confirm that the broad absorption has a simple Voigt line profile where the broadening is likely dominated by thermal Doppler broadening.

The Cosmic Origins Spectrograph (COS) provides a major new facility for studies of the low redshift IGM. Although COS has a spectral resolution several times lower than the E140M mode on the Space Telescope Imaging Spectrograph (STIS), COS is 10-20 times more efficient in collecting photons, making it possible to obtain relatively high S/N spectra of bright QSOs for absorption line studies of the IGM. With high S/N observations, the search for Ne VIII and/or broad H I absorption associated with O VI absorption is greatly facilitated. Therefore, we have initiated with COS a high S/N study of the physical conditions in IGM metal line systems with a goal of gaining a better understanding of the physical processes controlling the highly ionized absorbers found at low redshift.

The power of rest-frame far-UV and EUV observations for studying gas with T > $10^5$ K and even greater than T > $10^6$ K is not generally fully understood. Figure 1 displays ion column density versus log T for highly ionized species and H I assuming solar abundance ratios (Asplund et al. 2009) and collisional ionization equilibrium (CIE, from Gnat & Sternberg 2007) for a total hydrogen column density of logN(H) = 19. The CIE curves for O VI and Ne VIII peak at log T = 5.5 and 5.8, respectively. However, the CIE curves for these Li-like ions have broad tails that extend to log T ~ 6.0-6.5. The high temperature tails are the result of dielectronic recombination into the He-like ion where the recombining electron excites the core of the He-like ($1s^2$) electron configuration to produce a very short lived excited unbound Li-like intermediate state which radiatively rapidly decays to the bound Li-like ion in the electron configuration ($1s^2$ 2s). Thus two electrons participate in the recombination process. In the case of oxygen, the process produces a strong peak in the recombination cross-section at the ~0.5 keV energies corresponding to the 1s to 2p excitation energy of the He-like electron configuration of O VII.

Although the log N vs log T curve for O VI peaks in abundance at log T ~ 5.5, it is not necessarily true that most IGM O VI detections of collisionally ionized gas will trace gas with log T ~ 5.5. Gas in the temperature range from log T = 5.0 to 5.7 cools rapidly, while gas with log T ~ 6 cools slowly (Gnat & Sternberg 2007). If large column densities of gas with log T ~ 6 exist in the IGM, they could be detected via O VI , Ne VIII and broad H I absorption. For log T = 6 and log N(H) = 19 , we see from Figure 1 that log N(O VI) = 13.5, log N(O VII) = 15.7, and log N (H I) = 12.5 for Solar abundances. A 4σ UV detection of O VI with rest equivalent width, $w_r$ ~ 20 mÅ log N(O VI) ~ 13.3 is a relatively easy measurement for COS in spectra with S/N ~ 30 per resolution element. However, with current X-ray satellites, it is difficult to detect in absorption O VII columns smaller than log N(O VII) ~ 16.0 at a resolution of ~700 km $s^{-1}$. In fact, the only confirmed detections of O VII for intergalactic or circumgalactic gas are for gas at z = 0 associated with the Milky Way thick disk and corona where, for example, Fang et al. (2003) report log N(O VII) = 16.26 (+0.04, -0.25) for the line of sight to 3C 273 while Yao et al. (2008) find log N(O VII) = 16.00 (+0.16, -0.18) toward Mrk 421. In the case of H I, with S/N ~ 35 spectra from COS, it is possible to easily detect thermally broadened H I lines with b ~ 100 km $s^{-1}$ and log N(H I) > 13.5 provided the broad absorption is not confused with stronger overlapping H I absorption from cooler gas.

In this paper we report the detection of O VI and broad H I Lyman α (BLA) absorption in a multi-phase absorption system at z = 0.22601 toward HE 0153-4520 ($z_{em}$ = 0.450). The O VI and BLA measurements allow the determination of T, [O/H], and log N(H) in hot collisionally ionized gas. The observation illustrates the power of COS to detect and study gas in the low redshift IGM with T ~ $10^6$ K and a very large total column density of hydrogen. The simultaneous detection of both thermally broadened H I and O VI in the hot gas is required to determine the both the metallicity and total hydrogen content of the gas.

## 2. COS OBSERVATIONS OF HE 0153-4520

HE 0153-4520 with $z_{em}$ = 0.450 was discovered as part of the Hamburg/ESO survey for bright QSOs (Wisotzki et al. 2000). The QSO is bright in the UV with a 1120 to 1150 Å flux determined from FUSE to be ~2.6 x$10^{-14}$ erg $s^{-1}$ $cm^{-2}$ $Å^{-1}$. It is therefore well suited for high S/N UV absorption spectroscopy and was included in the COS GTO target list for the HST Program #11541, *Cool, Warm and Hot Gas in the Cosmic Web and Galaxy Halos*. The QSO is in the ROSAT bright object catalog with the ID 1RXS J015514.4-450605. The observed X-ray count rate of (8.5±1.6)x$10^{-2}$ counts $s^{-1}$ (Voges et al. 1999) is unfortunately ~90 times fainter than for the bright X-ray source 3C 273, so IGM absorption line X-ray studies of HE 0153-4520 are currently not possible.

HE 0153-4520 was observed by COS in December of 2009 (see Table 1). We have combined the COS G130M and G160M observations of HE 0153-4520 spanning the wavelength range from 1134 to 1796 Å. Information about COS can be found in Froning & Green (2009), and the COS HST Instrument Handbook (Dixon et al. 2010). The inflight performance of COS is discussed by Osterman et al. (2011) and in numerous instrument science reports found on the STScI COS website at http://www.stsci.edu/hst/cos/documents/isrs.

The spectral integrations were obtained with different grating set up wavelengths in order to obtain spectra with different detector/wavelength alignments to reduce the effects of detector fixed pattern noise and provide a way of covering the wavelength gaps between the two detector micro-channel plate segments (A and B). The various integrations and their HST MAST identification code, dates of observation, and integration times are given in Table 2.

The micro-channel plate delay line detector was operated in the time-tag mode with the QSO centered in the 2.5" diameter primary science aperture. The internal wavelength calibration lamps were flashed several times during each science integration. The time-tag data were processed with CalCOS with the reference files available 1 May 2010.

The detector and scattered light background levels in COS spectra are very small and unimportant. Flat-fielding, alignment and co-addition of the individual processed spectra utilized software developed by the COS team (release date 2/1/10) for the processing of FUV observations.[5]

The effects of major detector defects were removed when producing the combined spectrum by giving the affected wavelength regions in each individual spectrum low weight during the addition process . The reduced intensity in the grid wire shadows was also corrected for in the individual integrations and these affected wavelength regions were given lower weight in the co-addition process. The extraction process is explained in more detail in Danforth et al. (2010b).

The proper alignment of the individual spectra was achieved through a cross-correlation technique. The different individual spectra (in flux units) were weighted by integration time when combined.

The S/N per ~17 km $s^{-1}$ resolution element for the resulting combined spectrum spans the range from 20 to 40 for $1150 < \lambda < 1750$ Å with the peak S/N occurring in the 1400 to 1500 Å range. Since the weaker fixed pattern noise features were not flat fielded but simply averaged, there are residual features at the 2-3% level throughout the spectrum.

A characterization of the COS line spread function (LSF) is found in Ghavamian et al. (2009). The LSF has a narrow core and broad wings. At 1200, 1300, and 1400 Å the LSF has a full width at half maximum (FWHM) in velocity of 17.1, 15.4, and 13.9 km $s^{-1}$, respectively. However, the broad wings on the LSF have FWHM ~50 km $s^{-1}$ and contain ~20 to 30% of the LSF area. The broad wing contribution is largest at the shortest wavelengths. The COS G130M and G160M pixel sampling widths of 0.010 to 0.012 Å, provides ~ 6 to 8 samples per FWHM of the spread function. All COS spectra illustrated in this paper are shown with the full 0.010 to 0.012 Å sampling which corresponds to ~2.3 km $s^{-1}$ sampling.

Wavelengths, velocities and redshifts reported in this paper are heliocentric. The COS wavelength calibration is obtained through Pt lamp exposures obtained as part of each integration. Radio observations of H I 21 cm emission in the closest direction of HE 0153-4520 with the 36' beam of the Leiden-Argentina-Bonn survey (Kalberla et al. 2005) reveal two H I components of emission with $v_{LSR}$ = -7 and +8 km $s^{-1}$, FWHM = 17.0 and 12.5 km $s^{-1}$ and N(H I) = $(1.05\pm0.1)\times10^{20}$ and $(1.89\pm0.11)\times10^{19}$ $cm^{-2}$, respectively. Toward HE 0153-4520, $v_{HELIO} = v_{LSR} + 12.8$ km $s^{-1}$. We therefore would expect moderately strong ISM absorption lines of neutral gas tracers observed by COS to be centered near the $v_{HELIO}$ = 14 km $s^{-1}$ average velocity of the two H I components. Low ionization ISM lines observed in the G130M and G160M integrations were used to check the reliability of the COS velocity calibration. The ISM lines of P II λ1153, Si II λ1193, N I λλ1199.55, 1200.22, S II λλ1251, 1260, Ni II λ1371, Si II λ1527 yielded consistent average core heliocentric absorption velocities of 18±3 km $s^{-1}$ which is close to the H I 21 cm emission line reference velocity of 14 km $s^{-1}$. The COS wavelength calibration appears to be reliable from 1153 to 1527 Å. However, the ISM lines of Fe II λλ1144, 1608 and Al II λ1670 revealed substantial errors in the COS calibration at short ($\lambda <$ 1145 Å) and long ($\lambda >$ 1600 Å) wavelengths compared to wavelengths from 1153 to 1527 Å . It was therefore necessary to make velocity corrections to the absorption profiles for lines falling in the short and long wavelength ranges. The affected IGM lines in the system at z = 0.22601 include H I λλ 931, 926, and Si IV λλ1394, 1403. The H I lines were adjusted by +10 and + 28 km $s^{-1}$, respectively. The Si IV lines were adjusted by +22 km $s^{-1}$. The adjustments required by the ISM observations are consistent with those required to make the IGM system measurements internally consistent, After making adjustments to the wavelength scale at the short and long wavelengths we believe the resulting line profiles in the IGM system at z = 0.22601 have heliocentric velocities with ~ ±8 km $s^{-1}$ calibration errors which is ~½ a resolution element full width.

[5] See http://casa.colorado.edu/~danforth/costools.html for the COS team co-addition and flat-fielding software and for additional discussions.

## 3. FUSE OBSERVATIONS OF HE 0153-4520

A spectrum of HE 0153-4620 was obtained in 2003 by the Far Ultraviolet Spectrocopic Explorer (FUSE) satellite covering the wavelength range from 912 to 1184 Å with a resolution of 20 km $s^{-1}$. The 5602 second integration (MAST ID D8080301000 and program ID 808) in the 30"x30" square (LWRS) aperture produced a spectrum with S/N per 20 km $s^{-1}$ resolution element ~ 3 for λ > 1100 Å. Details about FUSE and its in orbit performance are found in Moos et al. (2000) and Sahnow et al. (2000). Because of the low S/N, the FUSE observations are not useful for the study of metal lines in the z = 0.22601 system, but the observations do permit a reliable estimate of the H I column density in the system from the Lyman limit absorption. The FUSE observations were calibrated using the CALFUSE version 3.2 (Dixon et al. 2007). The LiF2A and LiF1B observations were combined to produce the final spectrum from 1090 to 1184 Å, which is used for the Lyman limit absorption measurement discussed in §5.

## 4. COS OBSERVATIONS OF THE MULTI-PHASE ABSORBER AT z = 0.22601

Figure 2 shows the continuum normalized absorption line profiles as a function of velocity for the species detected in the system at z = 0.22601 along with several important non-detections. The QSO HE 153-4620 provided a very flat continuum making it easy to produce reliable continuum normalized absorption profiles. An example of the continuum placement is illustrated in Figure 3 in the 1487-1494 Å wavelength region containing the important redshifted H I λ1216 absorption line at 1490.4 Å. Species detected include H I λλ1216, 1026, 973, 950, 938, 931, 926, O VI λλ1032, 1038, Si IV λλ1394, 1403, C III λ977, N III λ990, Si III λ1207, C II λλ1036, 1335, N II λ1084, and Si II λ1260. N V λ1239 is only marginally detected. N V λ1243 and Si II λ1190 are not detected, while Si II λ1193 is contaminated by another IGM absorber at positive velocity.

The spectrum sampling shown in all the illustrations of COS spectra in this paper is ~2.3 km $s^{-1}$ while the spectral resolution (FWHM) ranges from ~ 14 to 17 km $s^{-1}$. Therefore the S/N per resolution element will be ~ 2.5 to 3 times larger than displayed in the 2.3 km $s^{-1}$ pixel samples illustrated in this paper.

The kinematical behavior of the absorption system is relatively simple. The low and moderate ions of C II, Si II, N II, C III, Si III, N III, and Si IV display relatively broad absorption lines extending from -40 to 40 km $s^{-1}$. C III, Si III and N III are the dominant ions in the gas containing the low and moderate ions and their profiles show no evidence for sub-structure. The profiles for the ions C II and N II possibly suggest sub-structure with components at approximately -10 and +10 km $s^{-1}$. However, that sub-structure is not evident in Si II λ1260 and Voigt profile fitting reveals that the substructure is not statistically significant (see below). The Si IV λ1394 line shows evidence for sub-structure but it is not evident in Si IV λ1403. If the sub-structure in these absorbers is real, its low contrast makes it very difficult to separate the absorption into multiple components. We will therefore treat the core absorption as a single absorption feature in the subsequent analysis.

The well detected lines of O VI exhibit broad symmetric absorption with no evidence for sub-structure. The H I absorption reveals a strong broad core in the all the Lyman series lines detected up to H I λ926. The very strong H I λ1216 line also reveals very broad H I absorption extending from -130 to +150 km $s^{-1}$ which appears shifted by ~15 km $s^{-1}$ with respect to the absorption in the narrower core component. We show in §6 that this broad H I λ1216 absorption is not the result of natural damping wings associated with the bulk of the H I absorption in the core component but is instead tracing thermally broadened H I in hot gas at logT ~6.

Comparisons of the apparent column density line profiles, $N_a(v)$, using the apparent optical depth method of Savage & Sembach (1991) for O VI λλ1032, 1038, Si IV λλ1394, 1403, and C II λλ1335, 1036 are shown in Figure 4. The good agreement for the separate $N_a(v)$ curves for the O VI and Si IV doublets with λf differing by a factor of two from the weak to the strong component implies the effects of line saturation are small for these two species. The good agreement for the C II λλ1335, 1036 lines also suggests little saturation although in this case the values of λf only differ by a factor of 1.34. The two $N_a(v)$ curves for C II illustrate the occasional ~8 km $s^{-1}$ velocity errors introduced by the COS wavelength calibration. The C III λ977 line is very strongly saturated while the Si III λ1207 line is strongly saturated.

The observed Lyman series lines are strongly saturated. Therefore, the H I column density in the strong core component is determined from FUSE observations of the Lyman limit break (see §5).

Basic measurements for the observed lines using the apparent optical depth (AOD) method of Savage & Sembach (1991) are given in Table 3 where for each line we list ion, rest wavelength, rest frame equivalent width in mÅ, average AOD velocity, $v_a$, the AOD Doppler parameter, $b_a$, the log of the apparent column density, $\log N_a$, and the velocity integration range of the measurements. The values of $b_a$ from the AOD method do not allow for instrumental blurring and only provide an approximate indication of the true line widths. Column densities are listed as lower limits if there is strong evidence for line saturation. Upper limits are 3σ limits. The notes to the table explain various issues affecting the measurements. Atomic parameters, including rest wavelengths and f-values, are from Morton (2003).

Voigt profile components fits to the observations were performed using the profile fit code of Fitzpatrick & Spitzer (1994). The code incorporated the wavelength dependent COS line spread functions derived by Ghavamian et al. (2009). The profile fit results are listed in Table 4 where the different columns list: absorption lines analyzed, rest wavelength(s), component velocity, v, component Doppler parameter, b, log N, the fit velocity range, and the value of the reduced $\chi_v^2$ for the fit. The notes provide information on various aspects of the profile fit results. For the metal line absorption, a single component adequately described the observations even though there is weak evidence for substructure in the C II and Si IV profiles. For H I a two-component fit was required to fit the strong core of the absorption and the broad wings on the H I λ1216 line (see §6). The simultaneous profile fit of O VI λλ1032, 1038 is shown in Figure 5. Additional profile fit examples for Si IV λλ1394, 1403, Si III λ1207, C III λ977, and C II λλ1036, 1335 are displayed in Figure 6.

Table 5 lists our final adopted absorber properties. We generally prefer the profile fit results since the fit process approximately allows for line saturation. However, the errors on the profile fit results are doubled to allow for systematic uncertainties associated with our lack of knowledge of the true kinematic properties of the absorption.

In the case of Si III λ1207, the profile fit column density is 0.13 dex larger than the AOD value of log $N_a$. This saturation correction will not be correct if the actual velocity structure in the Si III absorption is more complex than we have assumed. The unsaturated N III absorption is well fitted with a single absorption component with b = 24 km $s^{-1}$. The single component fit to the saturated Si III absorption yields b = 29 km $s^{-1}$ and log N(Si III) = 13.55. The larger b value for Si III compared to N III does not make sense if Si III and N III arise in the same gas. However, the larger inferred b value for Si III could easily arise if the absorption profile is somewhat more complex than we have assumed. If we fix the b value for the Si III absorption to be 24 km $s^{-1}$ and derive the value of the other parameters using the profile fit code, we obtain v = -4±1 km $s^{-1}$ and log N(Si III) = 13.68±0.02 and $\chi_\nu^2 = 1.07$. This exercise shows that the error in the Si III column density is very large. We therefore adopt the AOD column density log N(Si III) = 13.42 as a lower limit in Table 4 but note that the true column density could be 0.3 to 0.6 dex larger.

The C III absorption is so highly saturated we only give a lower limit to log N(C III) > 14.00 from the AOD method. However, we do note that log N(C III) = 14.64, 14.48, and 14.37 if b = 24, 26, and 28 km $s^{-1}$, respectively. Therefore, the true column density is probably substantially larger than the adopted lower limit.

The strongly saturated lines of C III and Si III exhibit weak absorption wings approximately 40 to 70 km $s^{-1}$ to each side of line center. The single component fits to these absorbers shown in Figure 6 reveals that these wings are the result of the broad component in the COS instrumental spread function.

## 5. THE H I COLUMN DENSITY FROM THE LYMAN LIMIT ABSORPTION OBSERVED BY FUSE

In Figure 7 we show the FUSE spectrum of HE 0153-4520 extending from 1094 to 1178 Å. In order to have the best available S/N, we used the combined LiF2A and LiF1B data. However, overplotting the LiF2A and LiF1B spectra shows the worm effect (see Dixon et al. 2007) decreases the flux in the LiF1B spectrum in the wavelength range 1151.9≤ λ ≤ 1165.6 Å, which is highlighted by the light line in the spectrum shown in Fig. 7. The Lyman series absorption lines in the z = 0.22601 system are marked and the Lyman series in the z = 0.22601 system converges to a Lyman limit break near 1118 Å marked by the dotted line. In order to estimate the Lyman limit optical depth in the Lyman Limit, $\tau_{LL}$, we need to model the QSO continuum before and after the break. The spectral resolution is unimportant for measuring the strength of the continuum break so we can heavily rebin the spectrum for our measurement. The bin size displayed in Fig. 7 is about 66 km $s^{-1}$, and we note that several bin sizes were tested with little effects on the determination of $\tau_{LL}$ and its error.

As the QSO exhibits several broad emission lines in the COS wavelength region, we find it more reliable to model the QSO continuum using the composite QSO spectrum derived by Zheng et al. (1997) rather than fitting the continuum using a polynomial function. This is especially critical since the Ne VIII doublet emission lines of the QSO could be present near the break region. We employed an IDL routine tested on a large number of QSOs and LLS (Ribaudo et al. 2011) to define the QSO continuum in a relatively absorption free wavelength range of the spectrum (several regions were tested and within ~1σ error the results were consistent with those derived below). Our QSO composite continuum is shown by the red line in Fig. 7 fitted to wavelengths longword of the LL break. Note that the area of the spectrum from 1152 to 1166 Å affected by the worm is not surprisingly poorly fitted. The blue curve shows the composite spectrum fitted to wavelengths blueward of the Lyman limit break. The optical depth at the LL is then derived from $\tau_{LL} = \log (\langle F_r \rangle / \langle F_b \rangle)$ where $\langle F_r \rangle$ and $\langle F_b \rangle$ are the averages fluxes defined by the red and blue curves, respectively, at $1095 \leq \lambda \leq 1115$ Å. We find $\tau_{LL} = 0.25 \pm 0.08$. The error on $\tau_{LL}$ includes contributions from the statistical uncertainities in the fit of the composite QSO continuum spectrum both on the long wavelength and short wavelength sides of the Lyman limit break. Possible systematic errors in the composite QSO spectrum adopted in the fitting process are not included in the listed error. The systematic error in $\tau_{LL}$ across the Lyman Limit break is estimated to be ~ ±0.05

With $N(\text{H I}) = 1.57\text{x}10^{17}\ \tau_{LL}(\lambda)\ (912/\lambda)^3$, the observed value $\tau_{LL}$ implies log N(H I) = 16.61 (+0.12, -0.17). This column density is consistent with the column density limits of logN(H I) > 16.22 and >16.38 derived from the AOD analysis of the two weakest but saturated H I lines at 931 and 926 Å. The H I column density from the LL break is important for modeling the core of the H I Lyman series absorption in order to determine the properties of the broad H I Ly α absorption wing seen in Figures 2 and 3.

## 6. OBSERVED PROPERTIES OF THE BROAD LYMAN-α ABSORBER

The broad absorption wings on the H I λ1216 line shown in Figures 2 and 3 are extremely important for understanding the origin of the O VI in the absorber at z = 0.22601. To determine the properties of the broad H I absorption we used the Voigt fitting code of Fitzpatrick & Spitzer (1994) to fit the H I Lyman series observations of COS.

The results from Voigt profile fits to strongly saturated absorption lines are very sensitive to the assumed properties of the absorption. When we first saw the broad absorption wings on the H I λ1216 absorption line, we believed we might be detecting the damping wings associated with a high column density in the core component to the H I absorption. However, the velocity offset between the core absorption and the wings was a potential problem. In first exploring the nature of the H I absorption we fitted a single component Voigt profile to the absorption observed in H I λ1216 and obtained the results: v = -10±1 km $s^{-1}$, b = 21.3± 0.2 km $s^{-1}$, log N(H I) = 17.87±0.02 for a fit over the velocity range from -320 to + 320 km $s^{-1}$ with $\chi_v^2 = 1.55$. However, a single component fit applied to the entire set of H I lines produced an acceptable fit to the narrow absorption observed in H I λλ1026 to 926 but failed to explain the broad wings on H I λ1216. Also, the large implied column density of log N(H I) = 17.87 obtained in the fit to H I λ1216 is completely inconsistent with the total H I column density obtained from the Lyman limit absorption observed by FUSE of log N(H I) = 16.61 (+0,12, -0.17). The small statistical errors obtained from the Voigt profile fit code given above are only correct if the assumed properties of the absorption are correct. In this case the absorption model is incorrect. The absorption is much better described by two components, one narrow and one broad. Assuming a single absorption component has produced a very misleading result. The broad wings on the H I λ1216 profile are not produced by H I damping wings. The broad wings must be from a second broad absorption superposed on the narrower absorption.

Two component Voigt profile fit results to the H I λλ1216 to 926 absorption shown in Fig. 8 are listed in Table 4. The error array for each H I line was determined from the observed noise in the continuum regions. The fit code allowed the velocities of the different H I lines to vary in order to correct for COS wavelength calibration errors. That procedure established that the observed H I λλ1216 to 938 lines have relative velocities good to 1-3 km $s^{-1}$ but that the H I λλ 931 and 926 lines required shifts of +10 and +28 km $s^{-1}$ as discussed in §2.

The effects of other IGM absorbers near the H I lines in the fit process were removed by making the errors at the wavelengths of those absorbers very large when performing the fitting. This gives the affected regions very low weight in the fitting process. The lighter gray lines in the spectra displayed in Fig. 8 show the regions of the spectra omitted in the fit process. Some of the omitted wavelength regions occur within the narrower core of the H I absorption lines. This includes the obvious contamination of H I λ973 from -120 to -70 km $s^{-1}$ and of H I λ938 from -130 to -70 km $s^{-1}$.

The H I column density in the strong saturated component is best constrained from the LLS absorption edge to be log N(H I) = 16.61 (+0.12, -0.17). Therefore, in order to measure the properties of the absorption in the broad H I component we specified the value of log N(H I) in the narrow component to be 16.61 and determined the value of all the other fit parameters for the two components as listed in Table 4. The best fit results for the two components fitted to the H I λλ1216 to 926 absorption obtained by adopting log N(H I) = 16.61 in the narrow component are:

Narrow component:

v = -11±1 km $s^{-1}$ , b = 27.0±0.1 km $s^{-1}$, and log N(H I) = 16.61 (fixed).

Broad component:

v = 5 ±6 km $s^{-1}$, b = 140±11 km $s^{-1}$, and log N(H I) = 13.70±0.03.

With $\chi_\nu^2 = 0.903$ the quality of the fit is good. The errors on v are only the statistical errors from the fit code and do not include the wavelength calibration errors. However, since the broad component is superposed on the narrow component of H I λ1216 absorption, the relative calibration error between the broad and narrow component should be small. We therefore conclude the BLA is shifted by 16±6 km $s^{-1}$ to positive velocity from the gas producing the narrow H I absorption. The listed error on b for the narrow component is not correct if the component structure of that absorber is more complex than the simple assumption of a single component with a Voigt profile. Errors for v , log N(H I), and b in the broad component introduced by the uncertainty in logN(HI) in the narrow component were determined by changing the value of log N(H I) in the narrow component over the acceptable ±1σ range from log N(H I) = 16.44 to 16.73. Through this process we find the ±1σ errors on v, b, and log N(H I) in the BLA from the uncertainty in the H I column density in the narrow component are v (+7, 0 km $s^{-1}$), b (+9, -11 km $s^{-1}$), and log N(H I) (+0.03, -0.07 dex).

The continuum in the region of the Ly λ1216 line as shown in Fig. 3 is very flat and well defined. However, to see if uncertainties in the continuum placement affect the values of the derived parameters for the BLA we increased and decreased the height of the continuum by acceptable amounts and re-derived the BLA properties for the two component absorption model. The ±1σ errors in the BLA fit parameters for v, b and log N(H I) from the continuum placement uncertainties are determined to be ±4 km $s^{-1}$, ±4 km $s^{-1}$, and ±0.02 dex, respectively. The continuum placement error is small compared to the statistical fit errors and the errors from the uncertainty in log N(H I) in the narrow component.

Combining the fit, log N(H I) and continuum placement errors in quadrature we obtain the following values of the parameters and their errors for the BLA: v = 5 (+10, -7) km $s^{-1}$, b = 140 (+14, -16) km $s^{-1}$, and log N(H I) = 13.70 (+0.05, -0.08) . The error on the column density implies the detection significance for the BLA is approximately 5σ.

We can not rule out the possibility that what we identify as a single BLA is instead two or more less broad absorbers. One absorber would need to be centered at ~ -100 km $s^{-1}$ and the other at ~ +100 km $s^{-1}$. Both these absorbers would need to have relatively large b values. There is no evidence for metal line absorption near these velocities and such an interpretation would require the O VI to be associated with the strong narrower H I absorption which appears to be ruled out by the photoionization modeling discussed in §7 and §8. As discussed in §9 and 10, the simplest and most logical explanation for the broad wings on the H I λ1216 line is the presence of a single BLA that is associated with the strong O VI absorption.

The BLA has a value of b = 140 (+14, -16) km $s^{-1}$. This is the broadest BLA so far reported for an IGM absorber with reliable measurements. Lehner et al. (2007) only list two BLAs with b > 100 km $s^{-1}$ with secure error estimates on the value of b. Both were measured by Sembach et al. (2004) in the spectrum of the bright QSO PG 1116+215. One is at z = 0.04125 with b = 105±18 km $s^{-1}$ and log N(H I) = 13.25±0.11. The other is at z = 0.9279 with b = 133±17 km $s^{-1}$ and log N(H I) = 13.39±0.09. Danforth et al. (2010) list both BLAs in their category of probable BLAs, but measure substantially smaller values of b. For weak BLAs the observed width of the feature is very sensitive to the continuum placement.

If the BLA at z = 0.22601 toward HE 0153-4520 has a line width dominated by thermal broadening, the very large Doppler parameter implies log T = 6.07 (+0.08, -0.10) We show in §8 that this BLA very likely is tracing the same hot gas that produces the strong O VI absorption.

## 7. PROPERTIES OF THE PHOTOIONIZED PLASMA IN THE PARTIAL LLS

In this section we investigate the properties of the plasma traced by the partial LLS with log N(H I) = 16.61 (0.12,-0.17). The species likely to exist in this plasma are H I, C III, Si III, Si IV, N III, CII, Si II, and N II. As we will see below, the O VI likely traces the absorption in much hotter gas. A first estimate for the temperature in the gas in the LLS follows from the measured H I b value from the profile fit listed in Table 4 of 27.0±0.1 km $s^{-1}$. If we take 27 km $s^{-1}$ as an upper limit to the H I Doppler parameter in the LLS we obtain log T < 4.6. Gas with log T < 4.6 is likely photoionized by the extragalactic background radiation field although we can not rule out contributions to the radiation from a source (galaxy) near the absorber.

Photoionization calculations were performed using Cloudy [ver.C08.00, Ferland et al. 1998] and assuming a uniform density one zone slab model illuminated by the extragalactic background radiation. We use the Haardt & Madau (2001) extragalactic background radiation field incorporating photons from AGNs and star forming galaxies but adjusted to that appropriate for the redshift, z = 0.226 of the absorption system. The assumed model structure is very simple. However, the observed nature of the absorption is also simple with the LLS lines being reasonably well modeled by a single absorbing component.

The heavy element reference abundances assumed in the photoionization model shown are for the solar photosphere from the new abundance compilation of Asplund et al. (2009) and reflect the recent changes in the solar abundances of N, C, and O.

The photoionization modeling results are shown in Figure 9 for a model with log N(HI) = 16.61 and [X/H] = -0.8 . The different curves show how the expected ionic column densities change with log U where $U = n_{\gamma}/ n_H$ is the ratio of ionizing photon density to total hydrogen (neutral +ionized) density. The heavy solid lines on the curves for each ion indicate the range of log U where the predicted column density agrees with the observed column density or its limit. The ion column densities are taken from the adopted values listed in Table 5. Figure 7 also includes scales for log (P/k) (K $cm^{-3}$), log T (K), log $n_H$ ($cm^{-3}$), log N(H), and log L(kpc), where H = H I + $H^+$ and L is the path length through the single zone absorber.

The best constraints for the ionization parameter in the photoionized gas come from those species with the most reliable column densities, particularly for adjacent ions of the same element. The column densities of N III and N II are well measured. The column density for Si II and Si IV are well measured but the Si III line is strongly saturated and we report a lower limit. The column density for C II is well measured but C III is so highly saturated we only report a lower limit to logN(C III).

If N III and N II exist in the same gas, their column densities imply log U = -2.8±0.1 and [N/H] = -0.8 (+0.3, -0.2). The relatively large errors on logN(H I) = 16.61 (+0.12, -0.17) contribute to the abundance error. For log U = -2.8 ±0.1 the photoionization model yields column densities for Si II, Si III, C II and C III consistent with the observations if [Si/H] = -0.8 (+0.3, -0.2) and [C/H] = -0.7 (+0.3, -0.2). Note that a 0.1 dex increase in the value of [C/H] from -0.8 to -0.7 dex is required to fit the C II observations in Figure 9 at log U = -2.8. Ignoring the 0.1 dex abundance difference for carbon, the once and twice ionized ions have column densities that are well described by by the photoionization model with log U = -2.8±0.1 and [X/H] = -0.8 (+0.3, -0.2).

The LLS absorber is almost certainly more complex than the simple one zone model assumed in the ionization calculation. This probably explains why the column densities for the once and twice ionized ions can be fitted at log U = -2.8±0.1 while the column density of Si IV requires log U = -2.5±0.1. The larger line width of Si IV (b = 34±1 km $s^{-1}$) versus N III (b = 24±1 km $s^{-1}$) also implies a more complex multi-zone absorber. The larger value of log U for Si IV could be achieved with a modest 0.3 dex decrease in density through the absorbing structure.

The abundances obtained for the partial LLS are [X/H] = -0.8 (+0.3, -0.2) for N, Si and C. Changing the shape of the assumed background ionizing radiation field can have a large effect on abundances derived from simple photoionization modeling as emphasized for another absorber by Howk et al. (2009). To illustrate the problem, we modified the photoionization model illustrated in Figure 9 by changing the ionizing background from the integrated AGN and star forming galaxy background adopted to the pure integrated AGN background of Haardt & Madau (2001). This simple change resulted in a 0.25 dex increase in the inferred value of [X/H].

Radiation from galaxies possibly associated with the absorber might also affect the intensity of the radiation field ionizing the absorber. However, without knowing the possible galaxy types and impact parameters it is difficult to speculate on how important the modifications might be except in the most general terms. The study of Narayanan et al. (2010) of an O VI absorption system at z = 0.226 toward H1821+643 considered the relative effects of the radiation from the extragalactic background and a $2L_*$ galaxy with an impact parameter of 100 kpc. The spectral shape of the galaxy spectrum was determined from the Milky Way high velocity cloud studies of Bland-Hawthorn & Maloney (1999) and Fox et al. (2005). The Milky Way model spectrum allowed 6% of the stellar ionizing radiation to escape into the halo. At z = 0.226 the ionizing radiation at 912Å from the $2L_*$ galaxy at 100 kpc was estimated to be ~6 times smaller than from the extragalactic background (see Fig. 5 in Narayanan et al. 2010). An impact parameter of 40 kpc would be required for the contribution from the $2L_*$ galaxy to be comparable to that from the extragalactic background. Since normal galaxies do not emit large fluxes of photons with E > 54 eV, the energy of the He II edge in hot stars, it is unlikely an associated galaxy will strongly influence the ionization of O VI unless the galaxy contains an active nucleus.

With the uncertainities in the extragalactic background radiation combined with possible additional ionizing radiation contributions from galaxies near the absorber we conclude, the derived abundances have systematic uncertainities that are difficult to quantify but could be as large as ~0.5 dex .

Adopting log U = -2.8±0.1, we obtain the following approximate parameters for the gas in the partial LLS model displayed in Figure 9: log N(H) = 19.35±0.18, log $n_H$ = -2.9±0.2, log T = 4.27±0.02, log (P/k) = 1.75±0.17, and log L(kpc) = 0.70±0.25. The model fit errors and errors on logN(H I) were added in quadrature. The expected thermal line width for the H I in the LLS is 17.6 km $s^{-1}$. The observed line width of b(H I) = 27.0 km $s^{-1}$ therefore implies a non-thermal contribution to the broadening of 20.5 km $s^{-1}$.

The partial LLS traces a structure where the plasma is relatively cool, confined, and at high pressure. The total hydrogen column density is log N(H) = 19.35±0.18.

## 8. CAN THE O VI BE PRODUCED BY PHOTOIONIZATION?

The observed Doppler parameter for the O VI line of 37±1 km $s^{-1}$ is much larger than the Doppler parameter for the H I in the LLS of 27.0±0.1 km $s^{-1}$. This implies it is unlikely the O VI exists in the cool photoionized gas producing the LLS. This conclusion is confirmed by the photoionization model displayed in Figure 9. At the value of log U = -2.8±0.1 that best explains the observations for the lower ionization absorption lines in the LLS, the predicted column density of O VI is log N(O VI ) = 10.5±0.5 while the observed column density of log N(O VI) = 14.21±0.02 is 3.7 dex larger. To explain the O VI via photoionization would require placing the O VI in a much lower density structure where the ionization parameter is much larger than log U = -2.8.

We note that the photoionization modeling of the once and twice ionized ions in the partial LLS at log U = -2.8±0.1 predicts logN(Si IV) = 13.0 while the observed value is logN(Si IV) = 13.62±0.04. The Si IV probably originates in a somewhat lower density zone associated with the photoionized absorber. The observed Si IV column density is produced if log U = - 2.5±0.1 implying the zone containing Si IV is a factor of two less dense than the zone containing the lower states of ioinization (see §7).

If Si IV and O VI are both produced by photoionization the curves of Figure 9 imply a value of log U = -1.6 is required to obtain log [N(O VI)/N(Si IV)] = 0.59±0.03. This requires a low very density and large path length of 1.8 Mpc over which the Hubble flow broadening would be several times more than the observed line widths. Such a model appears invalid because it also predicts N V and N III column densities ~ 1 dex larger than observed.

A photoionization model for the origin of O VI would leave unexplained the strong BLA with b = 140(+14, -16) km $s^{-1}$ and log N(H I) = 13.70(+0.05, -0.08) that is well aligned with the O VI absorption. The most logical explanation for the O VI absorption is that it occurs in the gas traced by the BLA. The properties of that gas are discussed in the following section.

## 9. PROPERTIES OF THE HIGHLY IONIZED HOT PLASMA TRACED BY O VI AND THE BLA

We showed in §8 that it is very unlikely that the strong O VI absorption occurs in the same gas producing the photoionized absorber containing C III, N III, Si III, Si IV, C II, N II and Si II and the strong narrow H I absorption. The O VI absorber is very well described by a single component Voigt profile with the following properties: v = -2 ± 1 km $s^{-1}$, b = 37±1 km $s^{-1}$ and log N(O VI) = 14.21±0.02. If thermal Doppler broadening dominates the broadening of the O VI absorption, the implied temperature is log T = 6.12±0.03 which is very similar to the temperature of log T = 6.07 (+0.08, -0.10) obtained from the BLA assuming its broadening is also dominated by thermal Doppler broadening. The BLA and O VI differ in velocity by 7 (+10, -7) km $s^{-1}$. The BLA and O VI therefore have the same velocity within the observational errors.

It is likely O VI and the BLA trace the same gas . Non-thermal contributions to the line broadening must be relatively small in order for two species that differ in mass by a factor of 16 to have b values that differ by a factor of 3.8±0.4. The temperature of the gas and the non-thermal contribution to the line broadening can be obtained by using b(O VI) = 37± 1 km $s^{-1}$ and b(H I) = 140(+14, -16) km $s^{-1}$ and solving $b^2 = b_t^2 + b_{nt}^2$, with $b_t = 0.129\ (T/A)^{1/2}$ km $s^{-1}$. We obtain log T = 6.07 (+0.09, -0.12) and $b_{nt}$ = 12 (+10, -12) km $s^{-1}$.

We conclude that the O VI and BLA are tracing the same hot gas with a temperature near log T = 6.07. Gas at this high a temperature will be mostly collisionally ionized except for a possible mild augmentation to the ionization from the background radiation field discussed below. Gas at this high a temperature cools slowly. Therefore, a good approximation to its ionization conditions is to assume collisional ionization equilibrum (CIE). Using the calculations of Gnat & Sternberg (2007) from the web site of Orly Gnat, http://wise-obs.tau.ac.il/~orlyg/cooling/CIEion/ , we obtain at log T = 6.07 (+0.09, -0.12) under CIE conditions, log (N(H)/N(H I)) = 6.71 (+0.12, -0.15) and log (N(O)/N(O VI)) = 2.61 (+0.10, -0.22). With log N(H I) = 13.70 (+0.05, -0.08) , log N(O VI ) = 14.21±0.02, we obtain logN(O VI)/N(H I) = 0.51 (+0.08, -0.05). Using the solar oxygen abundance from Asplund et al. (2009) of log (O/H) = -3.31, we find a logarithmic oxygen abundance in the hot gas of [O/H] = -0.28 (+0.09, -0.08). The total hydrogen column density in the gas is log N(H) = 20.41 (+0.13, -0.17).

We have investigated possible modifications from the CIE results above caused by additional ionization from the extragalactic background radiation. We calculated a hybrid collisional and photoionization model using the extragalactic background radiation from AGNs and starburst galaxies from Haardt & Madau (2001) for z = 0.226. The temperature of the gas was fixed at log T = 6.07 and the value of log [N(O VI)/N(H I)] was determined for a single slab model with log N(H I) = 13.70 and [O/H] = -0.15. The value of log [N(O VI)/N(H I)] is very close to the CIE value for log U < -2.0. For log U = -1.0 which corresponds to log $n_H$ = -4.7, log [N(O VI)/N(H I)] is only -0.10 dex different from the CIE result. A density of log $n_H$ = -4.7 implies a path length of 4.2 Mpc in the slab absorber to produce log N(H) = 20.41. Over such a distance the Hubble flow broadening of the absorber will be ~8 times wider than the observed absorber. We conclude that the effects of photoionization by the extragalactic background are negligible in influencing the derived properties of hot gas in the absorber.

The O VI and BLA in the absorber at z = 0.22601 toward HE 0153-4520 are therefore tracing hot collisionally ionized gas with a temperature of log T = 6.07 (+0.09, 0.12), a total column density of hydrogen log N(H) = 20.41(+0.13, -0.17), and an abundance of [O/H] = -0.28 (+0.09, -0.08). The total column density of hydrogen in the hot gas exceeds that in the cool photoionized gas of the LLS by a factor of ~11.

## 10. ARE THERE OTHER EXPLANATIONS FOR THE O VI AND BLA ABSORPTION?

In §9 we have discussed the simplest possible explanation for the origin of the O VI and BLA absorption in the multiphase system at z = 0.22601. In §8 we showed that it is unlikely the O VI arises in photoionized gas. Are there other possible explanations for the O VI and the BLA? The BLA could be totally unrelated to the O VI absorption. This requires explaining the large width for the BLA without invoking thermal broadening and finding an independent explanation for the O VI involving gas significantly cooler than logT ~ 6.

The large width for the BLA b = 140 (+14, -16) km $s^{-1}$ might arise from the relatively symmetric collapse or expansion of a gaseous region where the line of sight penetrates through both sides of the collapsing or expanding region. In this case we have no information on the temperature of the gas in the BLA. We can't rule out such an explanation.

With this independent explanation for the BLA, we also need an independent explanation for the O VI. If the LLS is tracing some type of collapsing structure where the denser part of the collapsing region is traced by the LLS, it is possible the O VI could trace hotter cooling gas in the collapsing structure. If the temperature of the O VI is substantially less than log T ~ 6, most of the broadening of the O VI would need to be from non-thermal motions involving the collapse. Those non-thermal motions are required to produce a very symmetric Gaussian O VI line profile that is well described by a Voigt profile with b = 37 km $s^{-1}$. In this cooling gas explanation for the origin of O VI, non-equilibrium ionization becomes important because the gas will cool more rapidy than it recombines. Other ions such as C III , Si IV and Si III should also trace the cooling gas but C III and Si III have much narrower absorption profile than O VI implying these ions exist in different zones of the cooling structure. With a single line of sight through a structure it is very difficult to truly rule out alternate explanations for an observed set of absorption profiles. It is possible the O VI arises in a cooling gas structure. However, in this cooling structure we now need to explain why the non-thermally broadened O VI with b = 37 km $s^{-1}$ and the non-thermally broadened H I with b = 145 km $s^{-1}$ share the same velocity to within ~ 15 km $s^{-1}$.

A possible alternate origin for the O VI absorption is place the site of the O VI in interface regions between the cool gas of the LLS and a hot exterior medium traced by the BLA. In this case the expected velocity of the O VI would be close to the velocity of the LLS as is observed. The expected column density of O VI per interface is logN(O VI) ~ 12.5-13.0 as observed in the LISM (Savage & Lehner 2006) and predicted theoretically (Böhringer & Hartquist 1987; Borkowski et a. 1990; Slavin 1989; Gnat et al. 2010). Therefore, for the interface explanation to be correct, ~20 to 60 interfaces would be required to achieve the large observed O VI column density of log N(O VI) = 14.3. The origin of the O VI line width for the interface explanation would be from the turbulence in the gas since individual interfaces would be expected to produce O VI lines with b ~ 18 km $s^{-1}$ which occurs for log T ~ 5.5 in the interface. Therefore, the turbulent broadening would need to be ~ 31 km $s^{-1}$ to produce the observed O VI line width of b = 37 km $s^{-1}$. While the turbulent mixing of cool and hot gas might provide the large number of interface zones and produce a symmetric O VI absorption profile, such an explanation for the observed O VI absorption appears contrived.

After going through this exercise in trying to find alternate explanations for the O VI and BLA, it appears that the simplest possible explanation for the origin of the O VI and the BLA makes the most sense. The O VI and BLA likely occur in the same hot gaseous region where thermal broadening dominates the observed line profiles and log T ~ 6.1.

## 11. POSSIBLE PHYSICAL LOCATION OF THE ABSORBER

Since a deep redshift survey has not yet been performed for the galaxies in the field surrounding HE 0153-4520, we need to draw on the results of previous galaxy/absorber association studies when commenting on the object(s) the absorber at z = 0.22601 may be associated with. Studies specifically aimed at probing the association of O VI absorbers with galaxies are those of Stocke et al. (2006), Chen & Mulchaey (2009) and Wakker & Savage (2009). Stocke et al. (2006) studied the galaxies associated with 37 O VI detections with logN(O VI) > 13.2 with z < 0.15 by drawing on 1.07 million galaxy redshifts and positions from the literature. They found the O VI absorbers nearly always lie within 800 $h_{70}^{-1}$ kpc of galaxies in regions where the sampling is complete to $L_*$. Chen & Mulchaey (2009) surveyed the galaxies in the fields around 3 QSOs with a total of 13 O VI systems and conclude that the O VI absorbers do not probe the gaseous halos around massive, early-type galaxies but do probe halos around emission-line dominated star-forming galaxies. Wakker & Savage (2009) studied H I/O VI absorption toward nearby (z < 0.017) galaxies for which the existing galaxy redshift literature allows the study of absorber/galaxy associations to relatively faint luminosity levels. For z < 0.085 the galaxy sample is essentially complete down to L = $0.1L_*$. Their sample included 14 O VI systems. They found that all 14 of the O VI absorbers occurred within an impact parameter of 550 $h_{70}^{-1}$ kpc from a L > 0.25 $L_*$ galaxy. The evidence for the association of O VI absorbers with the circumgalactic (halo) environments of galaxies is strong. It is therefore likely that the strong O VI absorber at z = 0.22601 toward HE 0153-4520 is associated with gas somehow connected to a galaxy. Speculations on the nature of that connection would be much more meaningful after information is obtained on the properties of the galaxy(s) through a galaxy redshift survey in the direction of HE 0153-4520.

The absorber potentially traces a large mass of gas if the covering factor of O VI in the circumgalactic regions of galaxies is large as suggested by the absorber/galaxy study Wakker & Savage (2009). For example, if the absorber traces a spherical halo region with a radius of 300 kpc and an impact parameter of 100 kpc the implied average density of gas along the line of sight is $n_H \sim 1.5\text{x}10^{-4}$ $gm^{-3}$ and the structure's mass is $\sim 5\text{x}10^{11}$ $M_O$. In this case, the pressure in the gas is log P/k ~2.6 which is ~0.8 dex larger than the value log P/k ~1.75 found in the LLS.

The high temperature of the gas in the O VI and BLA absorber implies the direct detection of a hot metal enriched plasma in what is likely the halo surrounding a foreground galaxy. The high metallicity of the absorber, [O/H] = -0.28 (+0.09, -0.08), implies the halo has been enriched by outflows from the associated galaxy and is not a primordial hot halo. The cooler photoionized gas in the absorber containing the partial LLS could represent a condensation in the hot halo or the direct detection of the extended gaseous disk of the galaxy. However, such explanations are difficult to understand if the abundance estimates for the hot gas , with [O/H] = -0.28 (+0.09, -0.08), and cooler photoionized gas, with [X/H] = -0.8 (+0.3, -0.2), are correct. Possible large systematic uncertainties in the abundance estimate for the cooler photoionized gas because of the uncertain knowledge of the appropriate shape of the ionizing background spectrum might remove the abundance differences. If the abundance estimates are correct, it is possible the LLS is tracing a cooler structure in the hot halo such as a tidal tail from a low abundance satellite galaxy similar to the Magellanic Stream in the Milky Way halo. We note that the estimated pressure in the LLS of log P/k = 1.75±0.17 is similar to pressures found in Galactic high velocity clouds including the Magellanic Stream (Fox et al. 2005, 2010). However, the small velocity offset between the hot gas O VI+ BLA absorber and the cooler partial LLS implies the two absorbers are situated in a structure with very simple kinematics.

## 12. SUMMARY

Observations of the QSO HE 0153-4520 ($z_{em}$ = 0.450) with the Cosmic Origins Spectrograph (COS) from 1134 to 1796 Å with a resolution of ~17 km $s^{-1}$ and S/N ~ 20 to 20 are used to study a multi-phase partial Lyman Limit system (LLS) at z = 0.22601. The Lyman Limit absorption is measured in an archival FUSE spectrum. The COS observations are analyzed in order to determine the physical properties of the plasma in QSO absorption line systems containing highly ionized atoms including O VI.

1. FUSE observations of the strength of the Lyman limit yield a H I column density of logN(H I) = 16.61 (+0.12, -0.17).

2. The observed UV absorption lines of H I Ly $\alpha$ to Ly $\lambda$ 926, C III, C II, N III, N II, Si III, and Si II imply the existence of cool photoionized gas in the LLS. Using Cloudy we derive log U = -2.8±0.1 and obtain the following approximate parameters for the gas in the partial LLS: log N(H) = 19.35±0.18, log $n_H$ = -2.9±0.2, log T = 4.27±0.02, log (P/k) = 1.75±0.17, and log L(kpc) = 0.70±0.25. The abundances are [X/H] = -0.8±0.2 for N, Si and C, although the result is sensitive to the assumed shape of the ionizing background radiation field.

3.The O VI and associated broad Ly $\alpha$ absorption with logN(O VI) = 14.21±0.02, b(O VI) = 37±1 km $s^{-1}$ , v = -2± 1 km $s^{-1}$, log N(H I) = 13.70 (0.05, -0.08), b(H I) = 140 (+14, -16) km $s^{-1}$, and v = 5 (+10, -7) km $s^{-1}$ and b(H I)/b(O VI) = 3.8±0.4 imply the direct detection of thermally broadened absorption by hot gas with log T = 6.07 (+0.09, -0.12), [O/H] = -0.28 (+0.09, -0.08) and log N(H) = 20.41 (+0.13, -0.17).

4. The total column density of hydrogen in the hot gas exceeds that in the cooler photoionized gas by a factor of ~11.

5. The O VI and BLA absorber traces a hot metal enriched plasma that likely exists in the halo surrounding a foreground galaxy. The cooler photoionized gas in the absorber traced by the partial LLS could represent the gas in a tidal tail similar to the low metallicity gas found in the Magellanic stream or the detection of cool gas in the extended outer parts of an associated galaxy.

6. The very high sensitivity of UV absorption line spectroscopy to O VI and thermally broadened H I Ly $\alpha$ now permits the direct detection of hot gas with log T > 6. The observations also allow a determination of the abundance of O/H and the total hydrogen (H I + $H^+$) column density, N(H) , in low-redshift IGM absorption line systems. However, the clear detection and analysis of the BLA in the system reported here requires spectra with relatively high S/N of ~35 to 40 per resolution element. Measures of the BLA along with O VI are crucial for determining the temperature, metallicity and total baryonic content of the absorber.

Acknowledgements: We thank the many people involved with building COS and determining its performance characteristics. We thank J. Ribaudo for sharing his QSO/LLS continuum IDL program we used to determine the strength of the LL break. We thank Andrew Fox and the anonymous referee for providing important suggestions that helped to improve the manuscript. BDS and AN acknowledge funding support from NASA through the COS GTO contract to the University of Wisconsin-Madison through NASA grants NNX08AC146 and NAS5-98043 to the University of Colorado at Boulder.

TABLE 1
LOW RED-SHIFT ABSORPTION SYSTEMS CONTAINING GAS WITH T > $10^5$ K[a]

| QSO | z | log N(O VI) | log N(Ne VIII) | log N(H I) | log N(H) | logT | Ref.[b] |
|---|---|---|---|---|---|---|---|
| HE 0226-4110 | 0.20701 | 14.37±0.03 | 13.89±0.11 | ~13.69 | ~19.9 | 5.7 | 1 |
| PKS 0405-123 | 0.49510 | 14.57±0.05 | 14.02±0.06 | … | ~19.7 | 5.7 | 2 |
| PKS 0405-123 | 0.16710 | 13.90±0.03 | … | <13.1 | ~19.7 | ~6.0 | 3 |
| PKS 0405-123 | 0.16716 | 14.72±0.02 | … | … | … | ~5.3 | 3 |
| 3C 263 | 0.32566 | … | 13.68±0.13 | … | … | ~5.7 | 5 |
| MRK 290 | 0.01027 | 13.80±0.05 | … | 14.35±0.01 | 19.6 | 5.15 | 6 |
| MRC 2251-178 | 0.00755 | 13.87±0.26 | … | 13.43±0.05 | 18.9 | 5.28 | 7 |
| MRK 876 | 0.00312 | 13.18±0.21 | … | 14.27±0.08 | 20.2 | 5.52 | 7 |
| 3C 273 | 0.00334 | 13.44±0.07 | … | 13.40±0.29 | 18.9 | 5.29 | 8 |
| PG 1444+407 | 0.22032 | 13.94±0.07 | … | 13.65±0.05 | 19.7 | 5.59 | 8 |
| H 1821+643 | 0.12147 | 14.02±0.07 | … | 13.78±0.17 | 19.4 | 5.38 | 8 |
| H 1821+643 | 0.22496 | 14.30±0.02 | <14.2 | … | … | ~5.3 | 9 |
| PG 1259+593 | 0.31978 | 13.49±0.07 | <13.57 | 13.98±0.06 | 19.9 | 5.52 | 10 |
| PKS 0312-77 | 0.2028 | 14.95±0.05 | … | … | … | 5.5-6.0 | 11 |

[a]Low redshift absorption systems for which a careful analysis provides moderate to strong evidence of the presence of gas with log T > 5. In some cases the evidence follows from the detection of both O VI and Ne VIII where the Ne VIII is very difficult to explain by photoionization in the low density IGM. In other cases, the evidence follows from the detection of O VI and an associated broad Lyman alpha line where the inferred temperature follows from the much larger line width of H I compared to O VI. The values of log N(H) are usually determined from log N(H I) using the inferred temperature and assuming CIE to calculate log [N(H)/N(H I)]. In several cases the values follow from a more careful evaluation of the effects of non-equilibrium ionization in cooling gas.

[b]References: (1) Savage et al. 2005; (2) Narayanan et al. 2011; (3) Savage et al. 2010; (5) Narayanan et al. 2009; (6) Narayanan et al. 2010a; (7) Wakker & Savage 2010; (8)Tripp et al. 2008; (9) Narayanan et al. 2010b; (10) Richter et al. 2004, (11) Lehner et al. (2009).

Table 2. COS G130M and G160M Integrations of HE0153 -4520[a]

| MAST ID | Date[a] | Grating | $\lambda_c$ [Å] | $\lambda_{mim}$ [Å] | $\lambda_{max}$ [Å] | t [sec] |
|---|---|---|---|---|---|---|
| LB6806010 | 12/19/09 | G130M | 1300 | 1146 | 1440 | 1143 |
| LB6806020 | 12/19/09 | G130M | 1309 | 1156 | 1449 | 1143 |
| LB6806030 | 12/19/09 | G130M | 1318 | 1165 | 1458 | 1471 |
| LB6806040 | 12/19/09 | G130M | 1327 | 1175 | 1468 | 1471 |
| LB6806050 | 12/19/09 | G160M | 1589 | 1401 | 1761 | 1471 |
| LB6806060 | 12/19/09 | G160M | 1600 | 1412 | 1772 | 1471 |
| LB6808070 | 12/19/09 | G160M | 1611 | 1424 | 1784 | 1471 |
| LB6806080 | 12/19/09 | G160M | 1623 | 1436 | 1796 | 1471 |

[a]The table lists the MAST exposure ID number, the date of observation, the COS grating, the central set-up wavelength, the minimum and maximum wavelengths covered by the integration, and the exposure time.

Table 3. COS AOD Measurements for the Multi-phase Absorber at z = 0.22601[a]

| ion | $\lambda_r$ | $w_r$ (mÅ) | $v_a$ (km $s^{-1}$) | $b_a$ (km $s^{-1}$) | log $N_a$ (dex) | [-v, +v] (km $s^{-1}$) | Note |
|---|---|---|---|---|---|---|---|
| H I | 1216 | 680±7 | -9±2 | 70±2 | >14.54 | [-150, 150] | 1 |
| H I | 1026 | 468±7 | -9±2 | 60±3 | >15.22 | [-150, 150] | 1 |
| H I | 973 | 421±6 | -13±2 | 56±3 | >15.62 | [-150, 100] | 1, 2 |
| H I | 950 | 346±6 | -7±2 | 47±3 | >15.85 | [-100, 100] | 1 |
| H I | 938 | 325±8 | -14±2 | 49±3 | >16.08 | [-100, 100] | 1 |
| H I | 931 | 288±9 | -7±2 | 47±4 | >16.22 | [-100, 75] | 1, 3 |
| H I | 926 | 280±14 | -16±2 | 45±5 | >16.38 | [-100, 75] | 1, 3 |
| O VI | 1032 | 148±6 | 0±2 | 48±4 | 14.18±0.02 | [-100, 100] | |
| O VI | 1038 | 91±7 | 3±3 | 53±6 | 14.22±0.03 | [-100, 100] | |
| N V | 1239 | 35±10 | 0±11 | 75±9 | 13.25:±0.13 | [-100, 100] | 4 |
| N V | 1243 | <30 | … | … | <13.35 | [-100, 100] | 5 |
| Si IV | 1394 | 224±13 | -4±3 | 47±5 | 13.56±0.02 | [-150, 50] | 6 |
| Si IV | 1403 | 173±13 | -8±3 | 55±5 | 13.69±0.03 | [-150, 50] | 6 |
| C III | 977 | 286±6 | -7±2 | 40±3 | >14.00 | [-100, 100] | 1 |
| C II | 1036 | 108±6 | -6±2 | 33±5 | 14.09±0.02 | [-75, 75] | |
| C II | 1335 | 170±10 | -9±4 | 43±4 | 14.04±0.03 | [-75, 75] | |
| N III | 990 | 115±11 | 1±3 | 38±5 | 14.16±0.04 | [-75, 75] | |
| N II | 1084 | 18±4 | -16±2 | 17±5 | 13.25±0.10 | [-50, 10] | |
| Si III | 1207 | 294±6 | -4±2 | 38±5 | >13.42 | [-100, 100] | 1 |
| Si II | 1190 | <20 | … | … | <12.76 | [-75, 75] | 5 |
| Si II | 1260 | 77±8 | -5±3 | 44±11 | 12.78±0.04 | [-75, 75] | 7 |

[a] Except where noted, all the column densities, velocities and line widths are derived from the AOD method. The AOD line widths, $b_a$, are not corrected for the effects of instrumental blurring. The errors listed only include the statistical uncertainty. The wavelength calibration errors introduce an additional velocity error of ~8 km $s^{-1}$ (see §2). Fixed pattern noise can influence the results for the weaker features at about the 5 mÅ level in equivalent width.

Notes: (1) The absorption is strongly saturated. The measured value of log $N_a$ is reported as a lower limit. (2) H I λ973 is contaminated by another absorber on the negative velocity wing. (3) H I λ931 and H I λ926 lie near the edge of the G130M integrations where the wavelength calibration is uncertain. The observed absorption line velocities were shifted by +10 and +28 km $s^{-1}$, respectively, to correct for the wavelength calibration errors (see §2). (4) The N V λ1239 line is only marginally detected. (5) A 3σ upper limit for $W_r$ is reported. (6) The difference in the values of log $N_a$ for the two Si IV lines implies a modest amount of line saturation. Using the saturation correction method of Savage & Sembach (1991) we obtain log N(Si IV) = 13.82± 0.05. The observed velocities of the Si IV lines have been increased by +22 km $s^{-1}$ to correct for a wavelength calibration error at long wavelengths (see §2). (7) The Si II λ1193 line is contaminated and the measurements are omitted. The Si II column density is therefore determined from the Si II λ1260 line.

Table 4. COS Profile Fit Results for the z = 0.22601 Absorber[a]

| ion | $\lambda_r$ | v (km $s^{-1}$) | b (km $s^{-1}$) | log N (dex) | [-v, +v] (km $s^{-1}$) | $\chi_\nu^2$ | Note |
|---|---|---|---|---|---|---|---|
| H I | 1216 to 926 | -11±1 | 27.0±0.1 | = 16.61 | [-320, 320] | 0.903 | 1 |
| | | 5±6 | 140±11 | 13.70±0.03 | [-320, 320] | | |
| O VI | 1032, 1038 | -2±1 | 37±1 | 14.21±0.01 | [-200, 200] | 0.667 | 2 |
| Si IV | 1394, 1403 | 0±2 | 34±1 | 13.62±0.02 | [-150, 150] | 0.955 | 2, 3 |
| N III | 990 | -3±2 | 24±1 | 14.23±0.03 | [-150, 150] | 0.946 | 4 |
| Si III | 1207 | -4±1 | 29±1 | 13.55±0.02 | [-150, 150] | 0.773 | 5 |
| C III | 977 | -8±1 | 15±1 | 16.49±0.13 | [-150. 150] | 0.773 | 6 |
| C II | 1036, 1335 | -11±1 | 28±2 | 14.11±0.02 | [-150, 150] | 0.798 | 2 |

[a]The Voigt profle fit code of Fitzpatrick & Spitzer (1994) and the wavelength dependent COS line spread functions from Ghavamian et al. (2009) were used to obtain the component fit results listed in this table. The listed errors are the statistical error from the fit process. For v there is an additional velocity calibration uncertainty of ~±8 km $s^{-1}$. The values of b and logN will be affected if the actual absorption is more complex than the simple assumed model absorption.

Notes: (1) A simultaneous fit to H I λλ1216 to 926 absorption lines assuming two components to the H I absorption with the H I column density for the narrower photoionized absorber of set equal to log N(H I) = 16.61 from the FUSE LL break analysis. All the other parameters were derived from the fit code. (2) Simultaneous fit to the lines listed. (3) There is some evidence for sub-structure in the Si IV λ1394 profile but it is not observed in the Si IV λ1403 profile. The substructure may be affecting the b value derived for Si IV. A two component free fit to the Si IV λ1394 line yields: v = -29±3 km $s^{-1}$, b = 25±2 km $s^{-1}$, logN(Si IV) = 13.47±0.05 and v = 8±2 km $s^{-1}$, b = 8±3 km $s^{-1}$, logN(Si IV) = 12.91±0.15. The total column density in both components is logN(Si IV) = 13.58±0.06. The possible presence of the sub-structure does not affect the derived column density very much because the absorption is only slightly saturated. (4) The continuum for the N III line is uncertain because the line lies at 1213.5 Å in the wing of very strong Galactic H I λ1216 absorption. (5) The Si III absorption is strongly saturated. The column density obtained from the free profile fit listed here is likely too small because the derived b value is too large to be consistent with the observations of N III and H I. (6) The C III line is very strongly saturated. Therefore the profile fit column density is extremely uncertain.

TABLE 5
Adopted Column Densities for the Multi-Phase System at z = 0.22601

| Ion | Log N | Note[a] |
|---|---|---|
| H I (narrow) | 16.61 (+0.12, -0.17) | 1 |
| H I (broad) | 13.70 (+0.05, -0.08) | 2 |
| O VI | 14.21±0.02 | 3 |
| N V | 13.25:±0.13 | 4, 5 |
| N III | 14.23±0.06 | 3 |
| N II | 13.25±0.10 | 4 |
| C III | >14.00 | 6 |
| C II | 14.11±0.04 | 3 |
| Si IV | 13.62±0.04 | 3 |
| Si III | >13.42 | 7 |
| Si II | 12.78±0.04 | 4 |

[a]Notes: (1) Column density is from the LLS break observed in the FUSE observations analyzed in §6. (2) The BLA column density from profile fitting and its errors are discussed in §6. (3) Column density is from the Voigt profile fit result listed in Table 4. The logarithmic errors for the metal lines have been increased by a factor of 2 to allow for systematic uncertainties associated with the simple model assumptions for the profile shape. (4) Column density is from the AOD value listed in Table 3. (5) result for N V is based on a marginal 3.5σ detection. (6) The C III line is strongly saturated. The AOD lower limit to the column density is listed. The true C III column density is likely to be ~0.5 to 0.8 dex larger, (7) The Si III line is strongly saturated. We report as a lower limit the AOD result. The true Si III column density is likely to be ~0.3 to ~0.5 dex larger.

FIGURES

FIG. 1. Column densities versus log T for highly ionized ions and H I assuming solar abundance ratios and collisional ionization equilibrium with log N(H) = 19. For log T from 5.5 to 6.5 the amount of O VI is 2 to 2.5 dex less than for O VII. However, COS can easily detect O VI column densities of log N(O VI) > 13.5 in absorption lines with b ~ 35 to 60 km $s^{-1}$ at a resolution of ~17 km $s^{-1}$. With current X-ray satellites it is difficult to detect O VII columns smaller than log N(O VII) ~ 16.0 at a resolution of ~700 km $s^{-1}$. If O VI and broad H I are detected it is possible to determine the temperature, oxygen abundance and total column of hydrogen for the gas.

FIG. 2 a and b. COS observations of continuum normalized flux versus rest-frame heliocentric velocity for various ions in the z = 0.22601 absorber over the velocity range from -300 to 300 km $s^{-1}$. The species are identified in each panel. The absorption profiles are displayed with the COS detector sampling of ~2.3 km $s^{-1}$ while the instrumental resolution ranges from ~14 to 18 km $s^{-1}$.

FIG. 3. COS observations of H I λ1216 absorption in the z = 0.22601 system. Flux is plotted against rest-frame heliocentric velocity over the range ±1000 km $s^{-1}$. The adopted continuum is displayed along with the error array for the ~2.5 km $s^{-1}$ pixel sampling. The observed photon counting S/N per 17 km $s^{-1}$ resolution element in this wavelength region is ~ 37.

FIG. 4. $N_a(v)$ profiles are compared for O VI λλ1032, 1038, C II λ1036, 1335, and Si IV λ1394, 1403. The good agreement for lines differing in λf by a factor of 1.4 to 2 implies little or no saturation in the absorption lines displayed.

FIG.5. Simultaneous Voigt profile fits to the O VI λλ1032, 1038 absorption lines. The fit is excellent with $\chi_v^2 = 0.667$ , implying the O VI absorption is well described by a single component Voigt profile with b(O VI) = 37 ±1 km $s^{-1}$ , log N(O VI) = 14.21±0.01 and v = -2±1 km $s^{-1}$.

FIG. 6. Single component Voigt profile fits to Si IV λλ1394, 1403, Si III λ1207, C III λ977, and C II λλ1036, 1335 in the absorption system at z = 0.220601. The pairs of Si IV and C II lines were fitted simultaneously.

FIG. 7. The FUSE observation of HE 0153-4520 from 1094 to 1178 Å showing the partial Lyman Limit break near 1118 Å from the absorption system at z = 0.22601. The expected position of the break is indicated by the vertical dotted line. The red curve shows the composite QSO continuum fitted to the wavelength region from 1120 to 1178 Å. The blue curve shows the absorbed composite QSO continuum fitted to wavelengths from 1095 to 1115 Å. The size of the Lyman break implies the optical depth and column density summarized in the figure.

FIG. 8. Simultaneous Voigt profile fits to H I λλ1216 to 926 assuming two H I absorbing components. The H I column density in the narrow H I component was fixed

at log N = 16.61 obtained from the Lyman Limit break in the FUSE observations of HE 0153-4520. The fitted lines show the total absorption and the individual narrow and broad absorption components. Velocity regions ignored in the fitting process are displayed with the lighter portions of the spectra.

FIG. 9. A simple single slab photoionization model for the moderately ionized gas in the z =0.22601 absorber. The radiation background is from the Haardt & Madau (2001) model interpolated to z = 0.226 and includes ionizing photons from AGNs and star forming galaxies. log N(X) for [X/H] = -0.8 is plotted against the logarithm of the photoionization parameter log U. Heavy solid lines on the lighter curves for the various ions display the observed column density range. The relative heavy elemental abundances adopted in the model are from Asplund et al. (2009).

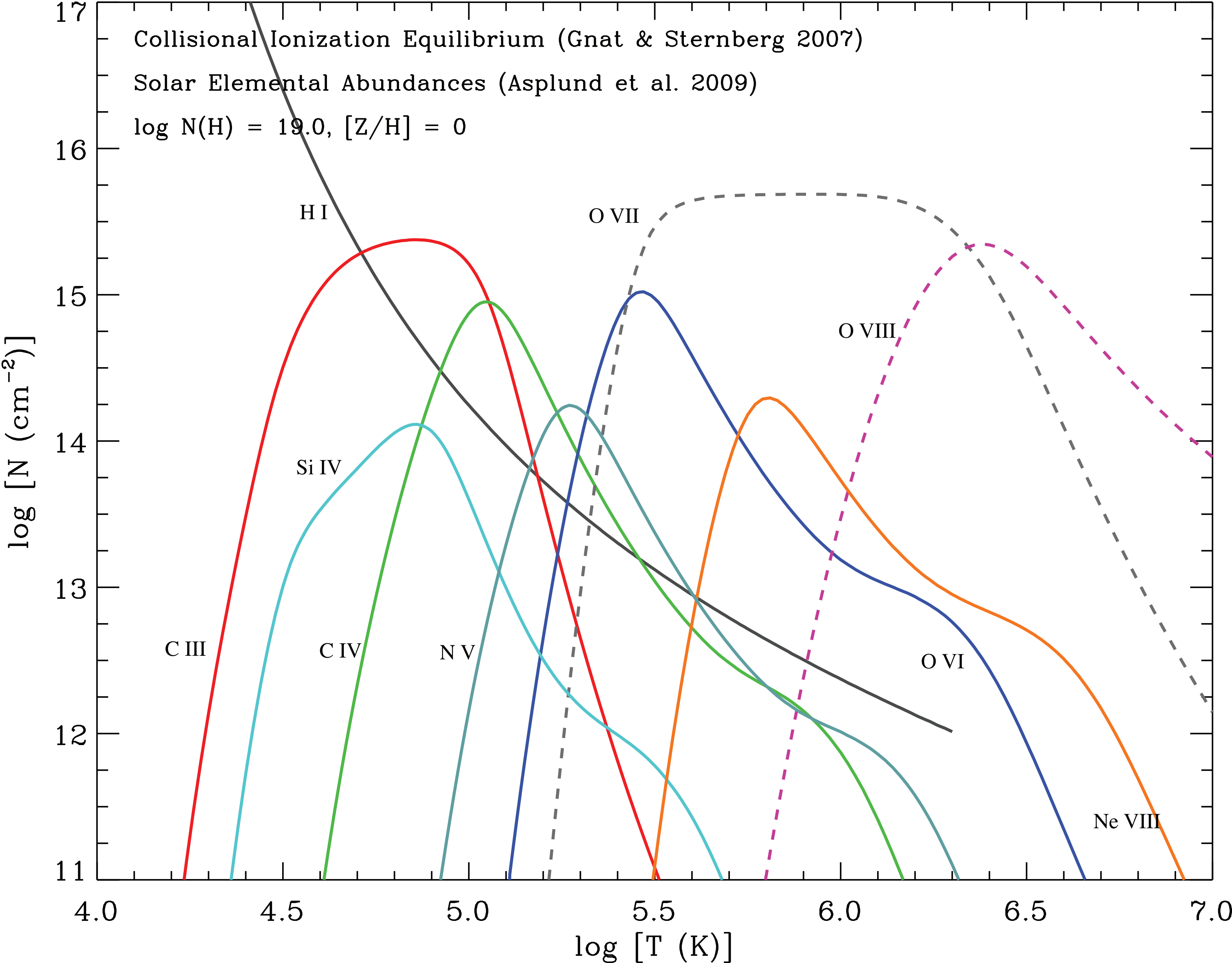

Collisional Ionization Equilibrium (Gnat & Sternberg 2007)
Solar Elemental Abundances (Asplund et al. 2009)
log N(H) = 19.0, [Z/H] = 0
H I
C III
Si IV
C IV
N V
O VII
O VIII
O VI
Ne VIII
log [N (cm$^{-2}$)]
log [T (K)]
17
16
15
14
13
12
11
4.0
4.5
5.0
5.5
6.0
6.5
7.0

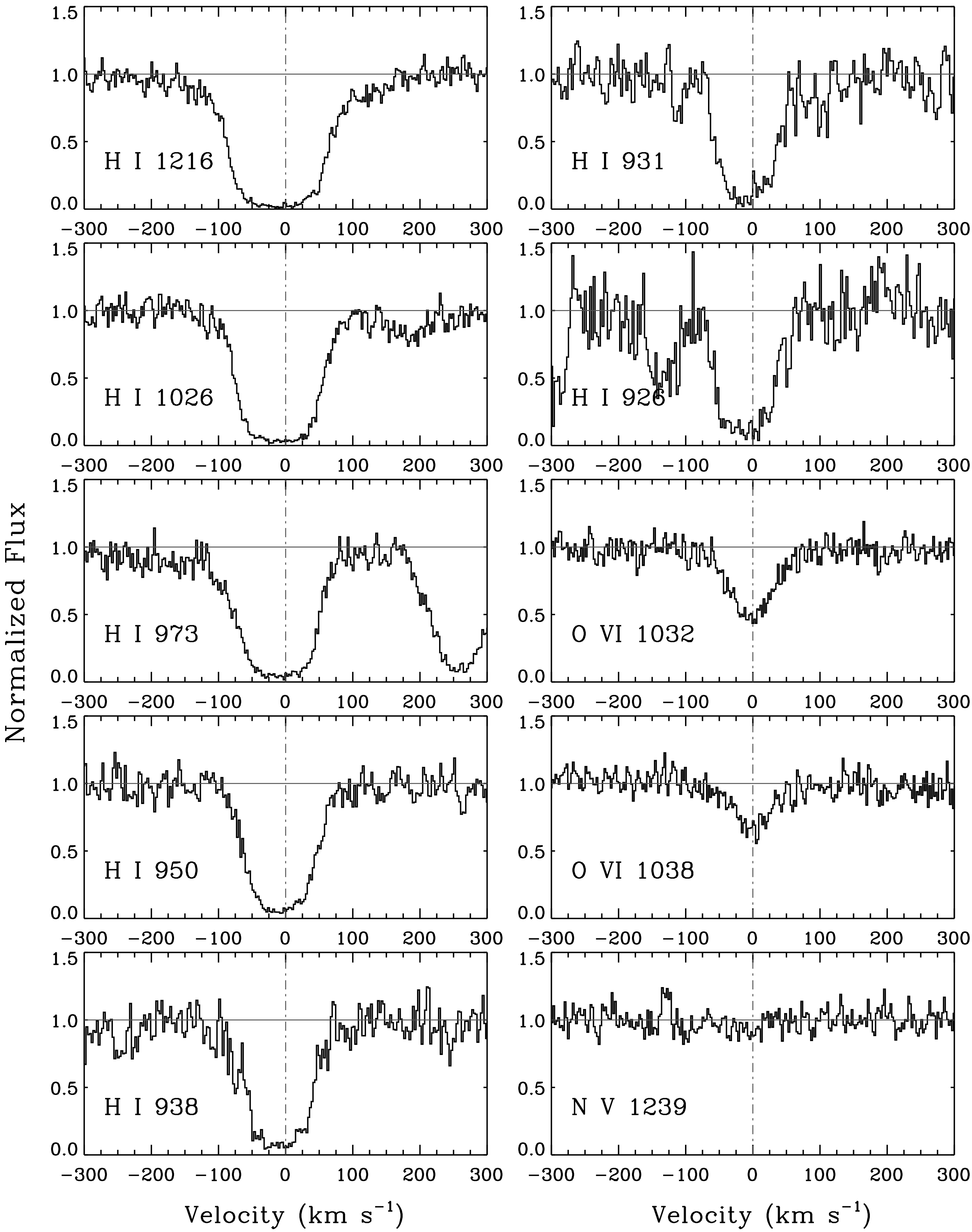
H I 1216
H I 931
H I 1026
H I 926
H I 973
O VI 1032
H I 950
O VI 1038
H I 938
N V 1239
Normalized Flux
Velocity (km s$^{-1}$)
Velocity (km s$^{-1}$)

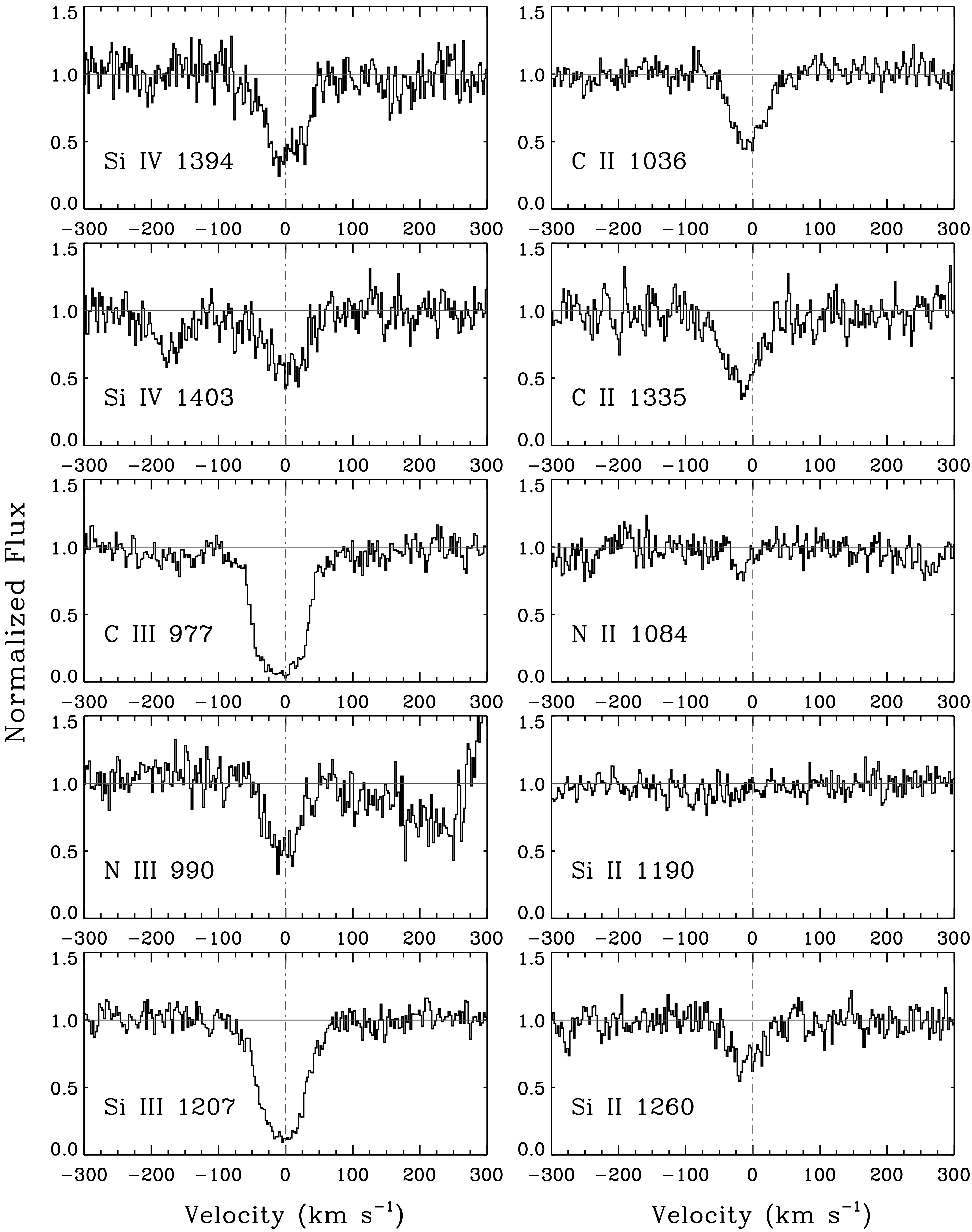
Si IV 1394
C II 1036
Si IV 1403
C II 1335
C III 977
N II 1084
N III 990
Si II 1190
Si III 1207
Si II 1260
Normalized Flux
Velocity (km s$^{-1}$)
Velocity (km s$^{-1}$)

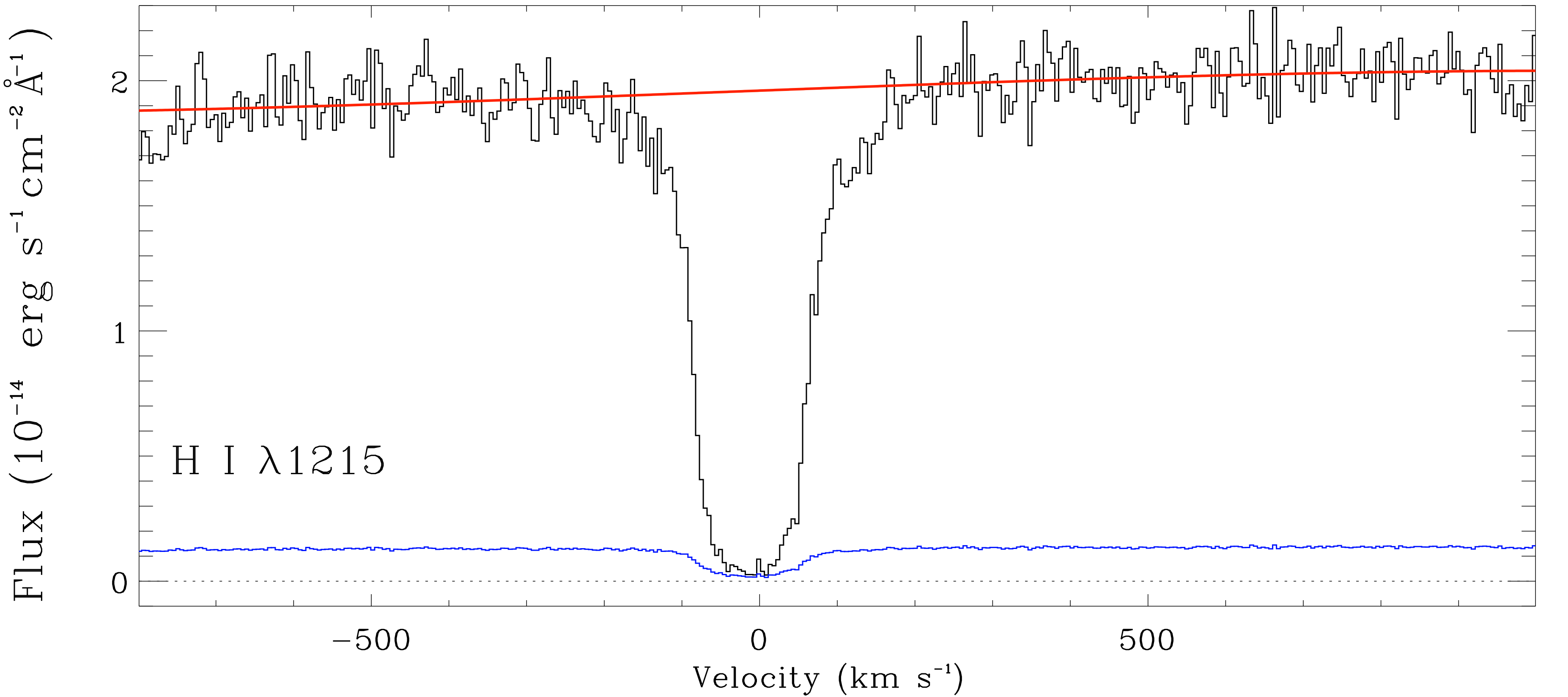

Flux ($10^{-14}$ erg $s^{-1}$ $cm^{-2}$ $Å^{-1}$)
2
1
0
H I λ1215
−500
0
500
Velocity (km $s^{-1}$)

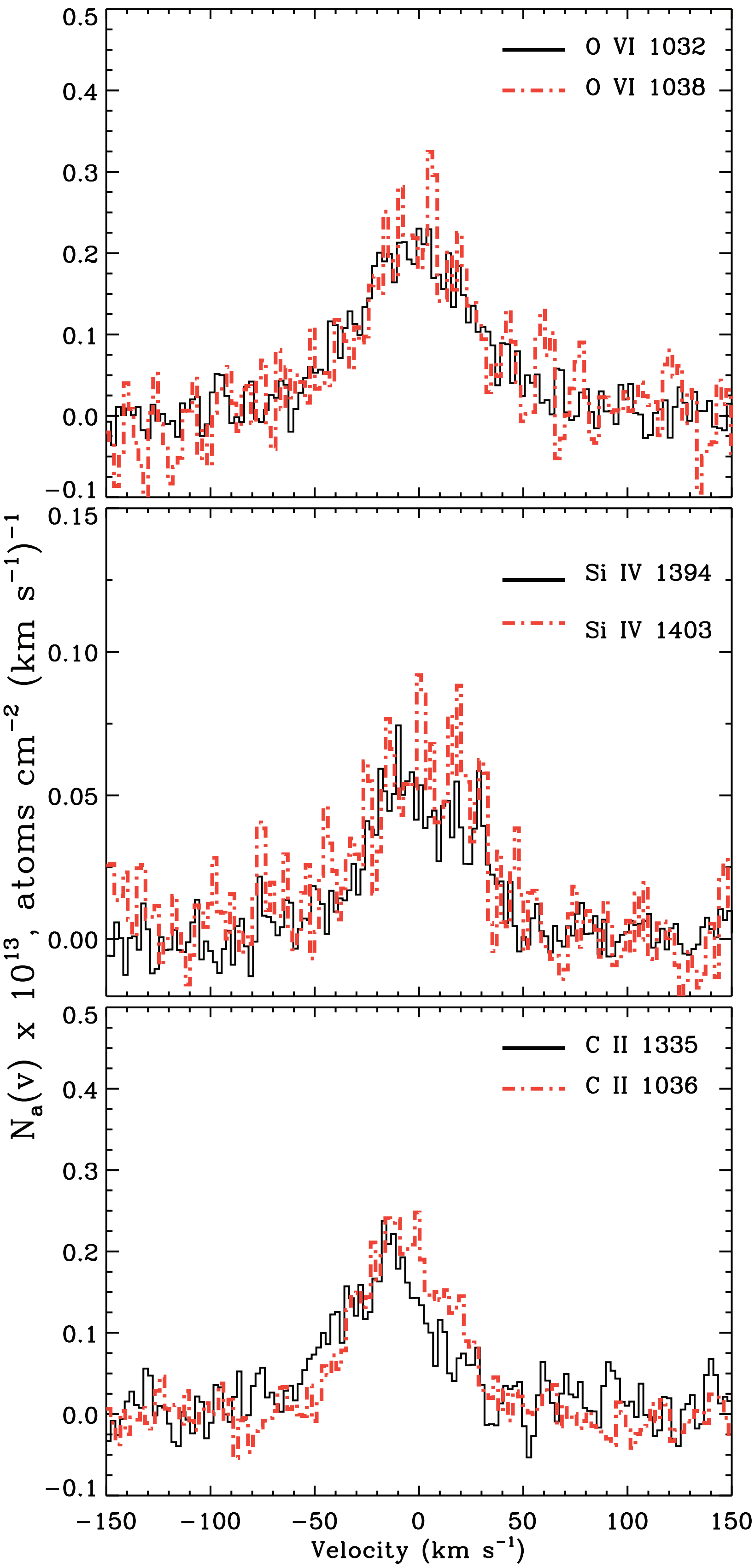

O VI 1032
O VI 1038
Si IV 1394
Si IV 1403
C II 1335
C II 1036
$N_a(v) \times 10^{13}$, atoms cm$^{-2}$ (km s$^{-1}$)$^{-1}$
Velocity (km s$^{-1}$)
0.5
0.4
0.3
0.2
0.1
0.0
−0.1
0.15
0.10
0.05
0.00
−150
−100
−50
0
50
100
150

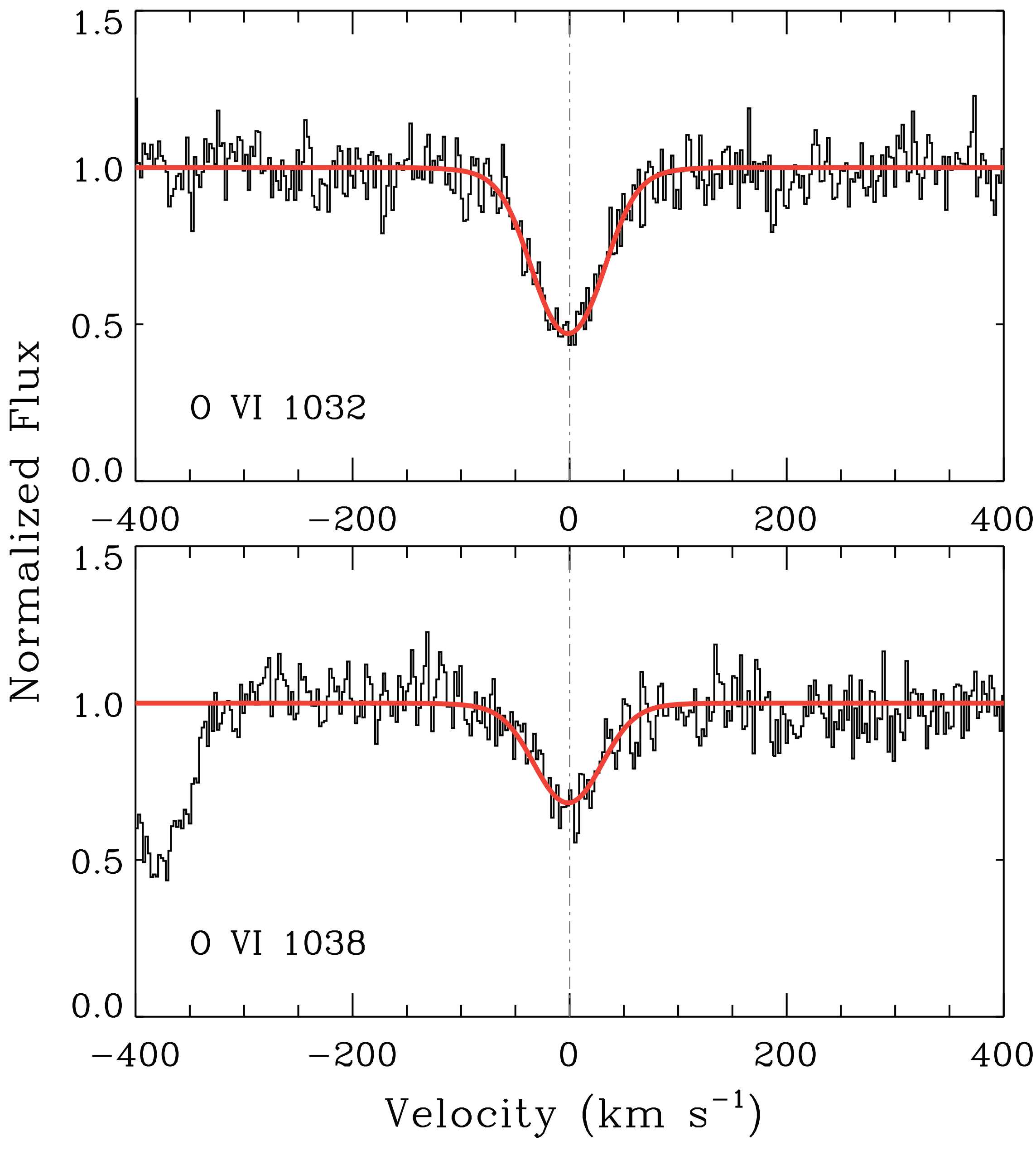

Normalized Flux
O VI 1032
O VI 1038
1.5
1.0
0.5
0.0
−400
−200
0
200
400
Velocity (km s$^{-1}$)

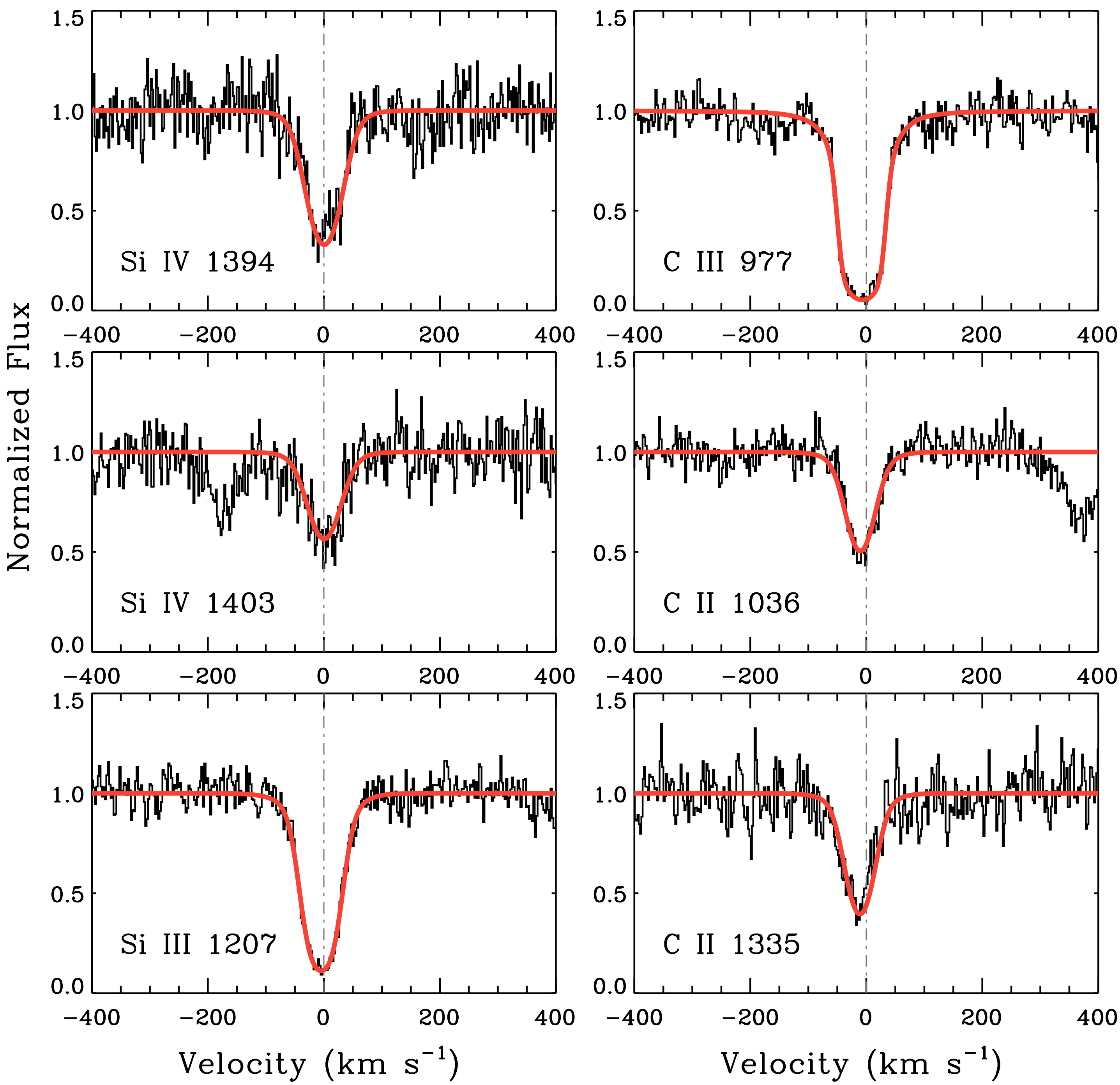

Normalized Flux
Si IV 1394
C III 977
Si IV 1403
C II 1036
Si III 1207
C II 1335
Velocity (km s$^{-1}$)
Velocity (km s$^{-1}$)

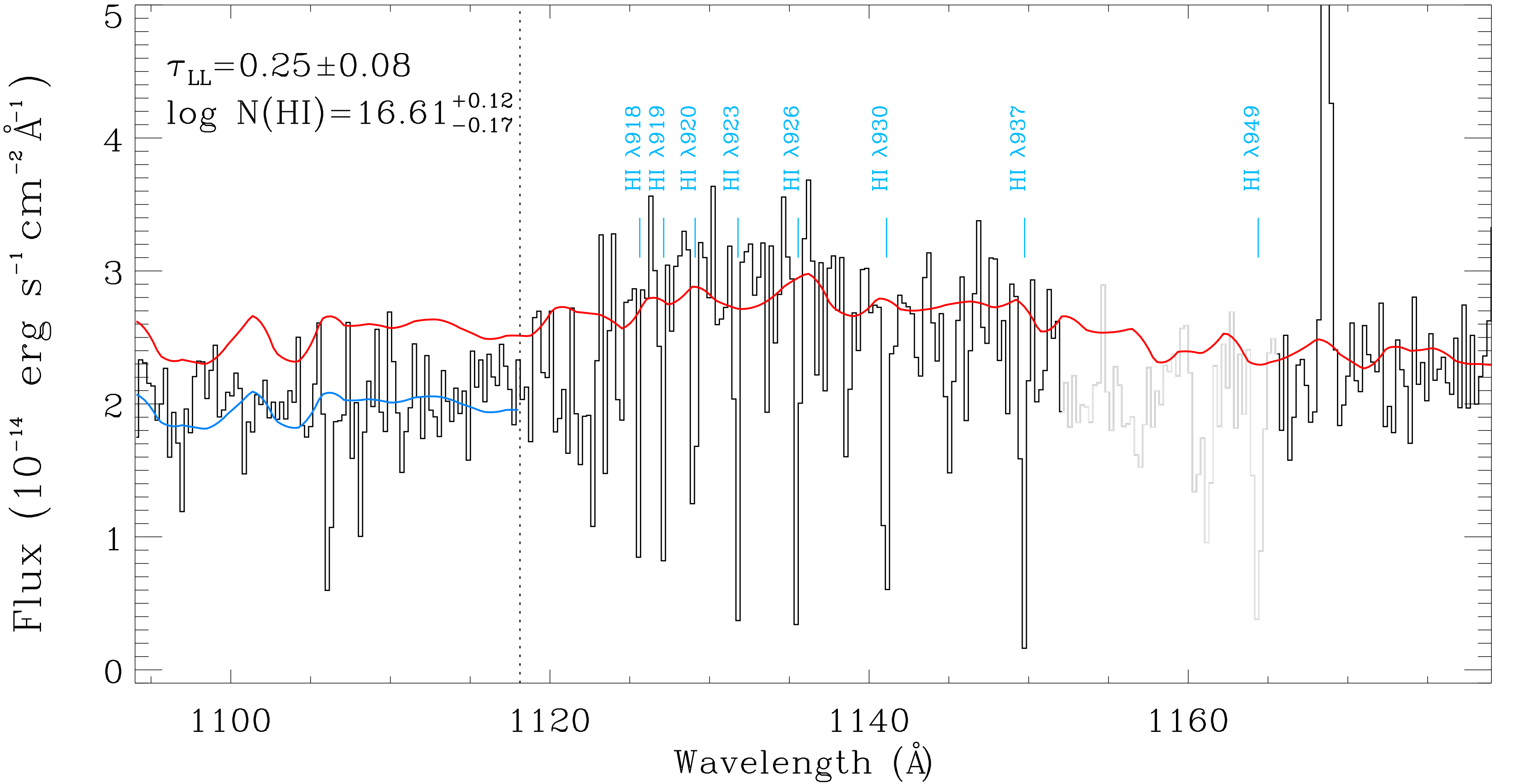

$\tau_{LL}=0.25\pm0.08$
log N(HI)$=16.61^{+0.12}_{-0.17}$
HI λ918
HI λ919
HI λ920
HI λ923
HI λ926
HI λ930
HI λ937
HI λ949
Flux ($10^{-14}$ erg s$^{-1}$ cm$^{-2}$ Å$^{-1}$)
Wavelength (Å)
0
1
2
3
4
5
1100
1120
1140
1160

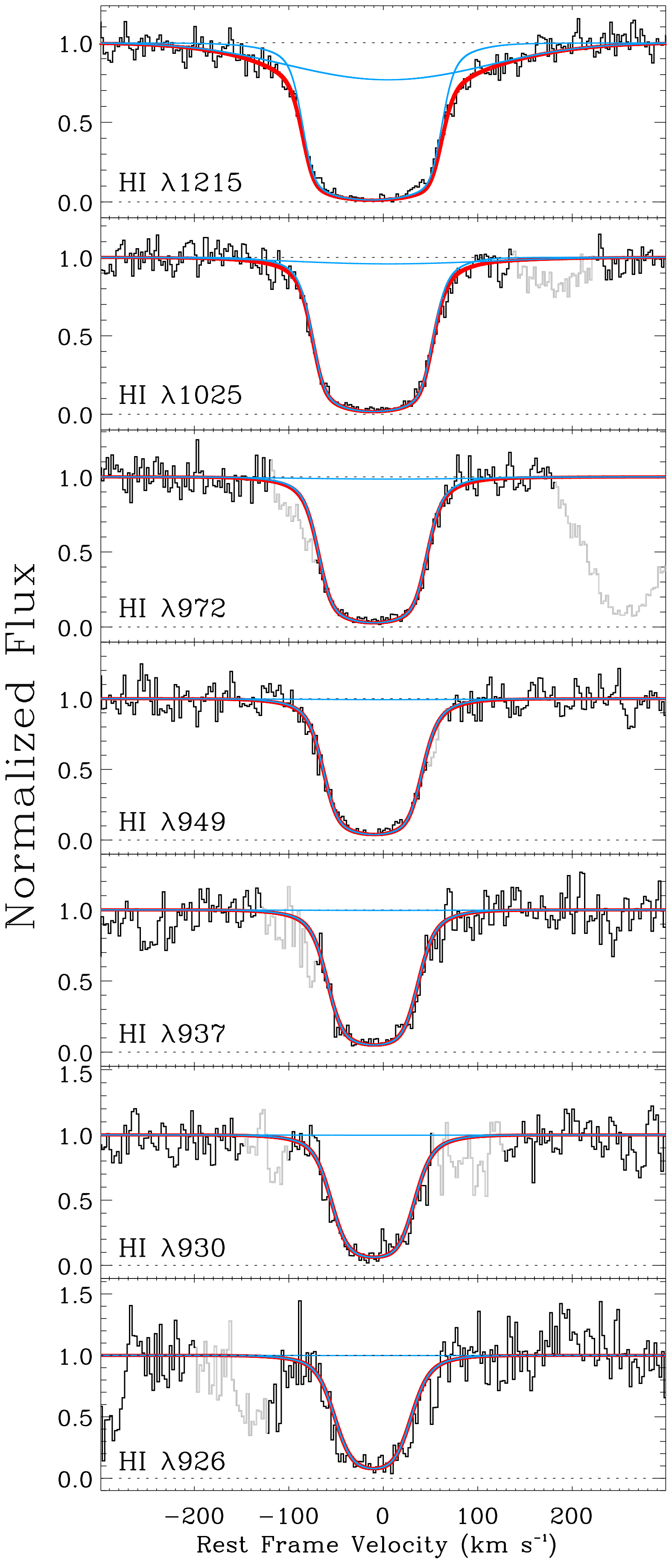
HI λ1215
HI λ1025
HI λ972
HI λ949
HI λ937
HI λ930
HI λ926
Normalized Flux
Rest Frame Velocity (km s⁻¹)
−200
−100
0
100
200
0.0
0.5
1.0
1.5

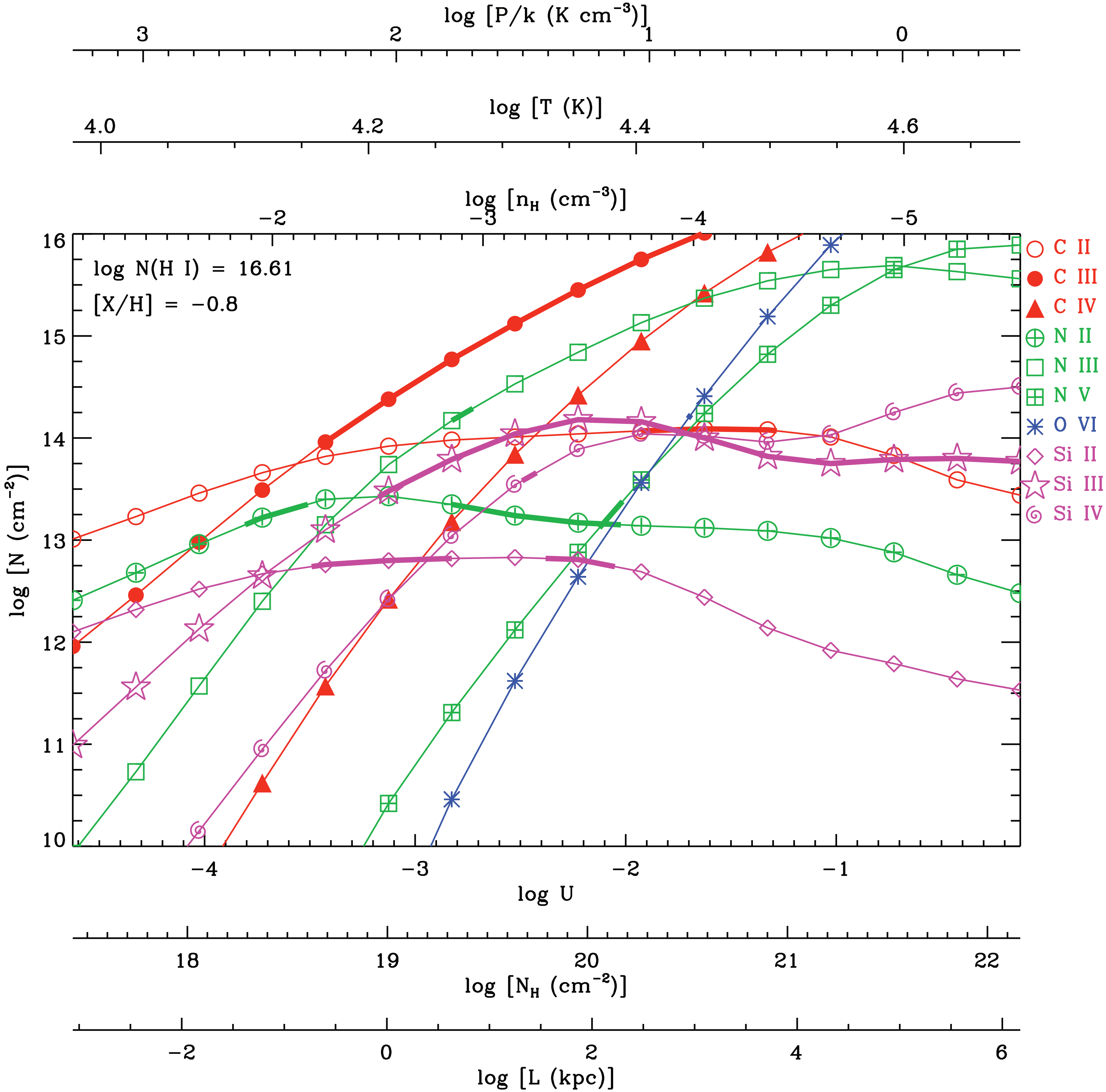

log [P/k (K cm^-3)]
3
2
1
0
log [T (K)]
4.0
4.2
4.4
4.6
log [n_H (cm^-3)]
−2
−3
−4
−5
log N(H I) = 16.61
[X/H] = −0.8
C II
C III
C IV
N II
N III
N V
O VI
Si II
Si III
Si IV
log [N (cm^-2)]
16
15
14
13
12
11
10
−4
−3
−2
−1
log U
18
19
20
21
22
log [N_H (cm^-2)]
−2
0
2
4
6
log [L (kpc)]